\documentclass[%
 reprint,
unsortedaddress,
 amsmath,amssymb,
 aps,
floatfix, 
]{revtex4-2}

\usepackage{graphicx}
\usepackage{dcolumn}
\usepackage{bm}

\DeclareUnicodeCharacter{2212}{-}

\usepackage[separate-uncertainty = true,multi-part-units=single]{siunitx}

\usepackage[colorlinks, linkcolor=black]{hyperref}

\usepackage{subcaption}%

\DeclareUnicodeCharacter{03C7}{$\chi$}

\begin{document}

\preprint{APS/123-QED}

\title{Positive Feedback Drives Sharp Swelling of Polymer Brushes near Saturation}

\author{Simon Schubotz}
\email{schubotz@ipfdd.de}
\affiliation
{Leibniz-Institut für Polymerforschung Dresden e.V, Hohe Stra\ss e 6, Dresden 01069, Germany}
\affiliation
{Technische Universität Dresden, Helmholtztra\ss e 10, Dresden 01062, Germany}

\author{Eva Bittrich}
\affiliation
{Leibniz-Institut für Polymerforschung Dresden e.V, Hohe Stra\ss e 6, Dresden 01069, Germany}

\author{Holger Merlitz}
\affiliation
{Leibniz-Institut für Polymerforschung Dresden e.V, Hohe Stra\ss e 6, Dresden 01069, Germany}

\author{Quinn A. Besford}
\affiliation
{Leibniz-Institut für Polymerforschung Dresden e.V, Hohe Stra\ss e 6, Dresden 01069, Germany}

\author{Petra Uhlmann}
\affiliation
{Leibniz-Institut für Polymerforschung Dresden e.V, Hohe Stra\ss e 6, Dresden 01069, Germany}

\author{Jens-Uwe Sommer}
\affiliation
{Leibniz-Institut für Polymerforschung Dresden e.V, Hohe Stra\ss e 6, Dresden 01069, Germany}
\affiliation
{Institute for Theoretical Physics, Technische Universität Dresden, 01069 Dresden, Germany}

\author{Günter K. Auernhammer}
\email{auernhammer@ipfdd.de}
\affiliation
{Leibniz-Institut für Polymerforschung Dresden e.V, Hohe Stra\ss e 6, Dresden 01069, Germany}

\date{\today}

\begin{abstract}
We resolve the Schr\"{o}der paradox for PNiPAAm brushes, showing experimentally that swelling at 100\% relative humidity (RH) matches the liquid state. This occurs via a sharp increase in swelling above 98\%~RH, a behavior standard models fail to explain. Our extended mean-field theory explains this via a positive feedback between swelling and solvent quality, driven by a concentration-dependent $\chi$ parameter. The swelling isotherm quantitatively predicts the dynamic wetting crossover: the advancing contact angle at high velocities drops sharply as ambient humidity surpasses the 98\%~RH threshold.
\end{abstract}

\maketitle

Polymers can respond to a variety of stimuli, such as pH \cite{W.1950, Karg2008}, temperature \cite{Taylor1975,Akiyama2004, Plamper2017}, light \cite{Kollarigowda2017,Ishihara1984, Groten2012} and many others \cite{Heskins1968, Stuart2010, E2004}. 
This sensitivity makes them ideal candidates for coatings to control processes like cell adhesion \cite{Akiyama2004} or antifouling \cite{Magin2010, Demirci:2017aa} and applications like  filtration \cite{Belfort1994,Zhou2018}, drug delivery \cite{Qiu2001}, sensing \cite{Islam2015, Schoenberg2016}, or others \cite{Chen:2010af, Poppinga2017, Galaev1999}. 
Thin polymer coatings can be produced by end-grafting polymer chains onto a surface \cite{Gennes1980, MILNER1991}, where the long polymer chains are grafted so densely that steric interactions forces the chains entropically away from the surface, referred to as a polymer brush.
Similarly to bulk polymers, polymer brushes swell when in contact with solvents  \cite{Brown_1990aa, WITTMER199485, Tsujii2006, 2004a}. 
From a theoretical standpoint, the driving force for the polymers' response is the chemical potential; thus, in a fully saturated gas phase, the response should be the same as in liquid \cite{Birshtein1994, Eck2022}. 
In 1903, Schr\"{o}der observed that gels swell more in liquid than in a ``completely'' saturated gas phase. Since then, a difference between swelling in liquid and swelling in a saturated gas phase is often observed \cite{CHEN2023113050}. However, these observations are also controversially discussed \cite{Onishi_2007aa, Beers_2014aa, JECK201174}. Various solutions have been proposed to resolve this controversy: possible heterogeneities in the material should be taken into account \cite{VALLIERES2006357, C1SM05273J}, strong changes near absolute saturation \cite{Beers_2014aa} make extrapolations tricky, and surface contributions to free energy \cite{Hartmann_2024aa} should be considered. 

Using stimuli-responsive polymers, like poly(N-isopropylacrylamide) (PNiPAAm) \cite{Kubota1990}, gives polymer brushes with pronounced responsiveness \cite{Bittrich2012, Plamper2017}. 
Most commonly, temperature is used as external trigger.
PNiPAAm chains in aqueous solution collapse above a critical temperature called the lower critical solution temperature (LCST). 
The LCST depends on the polymer concentration in solution \cite{Afroze2000, Halperin2015}.
This behavior can be modeled using the Flory-Huggins approach, with an interaction parameter varying with polymer concentration \cite{Afroze2000}.
Whereas the properties of such brushes in full contact with a liquid have been studied in detail, only recently, studies have been reported of partially wetted polymer brushes, i.e., of brushes wetted by single drops \cite{Butt2018, Hartmann_2024aa}. 
One and the same brush can have different wetting properties \cite{Schubotz2021, Sychev2023}, which can be controlled via the vapor phase \cite{Schubotz2023, Diekmann2024}, or the wetting history \cite{Schubotz2021}.
For partially wetted brushes, there is a gradual increase in vapor saturation to full saturation near the drop, leading to brush swelling around the drop, creating a halo \cite{Kap_2023aa, Besford2022, Besford2022a}.
Since at the free surface of a drop the surrounding atmosphere is fully saturated, the response of polymer coatings to the full range of relative humidities (RHs) is relevant. 
In studies performed in the vapor phase, the response was significantly less as full saturation was not reached \cite{Biesalski:2000aa, RitsemavanEck2022}.  
This behavior is an example of adaptive wetting, where the substrate's properties change in response to the liquid~\cite{Butt2018}.

In this Letter, we address the century-old Schr\"{o}der paradox by investigating the thermodynamic response of PNiPAAm brushes to a vapor phase. Using a novel setup for precise control of RH and with that chemical potential, we resolve the paradox by measuring the full swelling isotherm up to 100\%~relative humidity (RH). We find that swelling in a fully saturated vapor matches the liquid state but occurs via a sharp continuous increase concentrated above 98\%~RH. This behavior is not captured by standard mean-field theory. Because of this, we extend the mean-field model by a concentration-dependent $\chi$ parameter, determined directly from our experimental data, which reveals a positive feedback loop between swelling and solvent quality. This sharp increase in equilibrium swelling is directly mirrored in the dynamic wetting, creating a velocity-dependent crossover: at high speeds, the advancing contact angle is governed by ambient humidity, while at low speeds, it is dictated solely by the droplet's local vapor halo, which in close proximity to the contact line is nearly independent of the ambient.

We build on our methods described previously \cite{Schubotz2023}. 
For consistency, we give only a brief summary here; full experimental details are provided in the Supplemental Material (SM)~\cite{SM}, Sec. S1.

Polymer brushes were synthesized on silicon using the grafting-to method, resulting in a dry thickness of about \SI{10}{\nano\meter}. 
We determined the brush thickness with a ellipsometer, on which we use a custom-built cell that can be filled with liquid or vapor to vary conditions. 
The resulting data were determined by a box model using CompleteEASE software (J. A. Woollam Co. Inc.). 
To control the temperature when measuring in various vapor concentrations, the entire ellipsometer is placed in the climate chamber, the LHU-113 (ESPEC North America, Inc., in Hudsonville, MI, US) with a temperature accuracy of \SI{0.5}{\degreeCelsius}. 
For RHs in the range of \SIrange{10}{90}{\percent}, we used appropriate mixtures of dry and saturated nitrogen flux through the measurement cell. 

RH and temperature are measured with the HYT 939 sensor (ist AG, Switzerland), which operates from \SIrange{0(1.8)}{100(1.8)}{\%} RH and \SIrange{-40(0.2)}{125(0.2)}{\degreeCelsius}.
Data is read using the HYT module Evalkit (iST AG, Switzerland) and saved to a CSV file via a Python script. 
The sensor is integrated into the 3D-printed lid that seals the ellipsometer cell.
Data from all devices is merged and processed using an electronic lab book \cite{schubotz2025}. 

For RHs up to 100\%, we placed water-soaked tissues in the ellipsometry cell and remoistened them with a syringe before starting the experiment (see schematic in Fig. S1a in the SM~\cite{SM}).
The tissues offered a large surface area for water evaporation and occupied about half of the measuring cell.
Due to a small temperature gradient between the bulk of the ellipsometer cell and the sample surface, there was water condensation on the sample. 
We regulated the RH in the range of \SIrange{80}{100}{\percent} at the sample using a flow of dry nitrogen through the ellipsometer cell. 
The RH at the sample was determined by correcting the readings from the sensor in the cell lid for thermal lag and calibration drift, as detailed in Sec. S1 E in the SM~\cite{SM}.
To determine precisely the RH just above the sample, we use the onset of condensation on the sample; see below for details. 

For the contact angle measurements, we controlled humidity around the drop using a gas mixture from gas wash bottles, directed into our contact angle measurement cell \cite{Schubotz2023}. 
The target RH was achieved by actively cooling the sample with a Peltier element. 
A custom script continuously calculated the required sample temperature in real-time using a Magnus-type empirical formula~\cite{Alduchov1996} and data from a HYT 939 sensor. 
This target temperature was then sent as a setpoint to a TEC-1161-4A controller (Meerstetter Engineering GmbH), which regulated the power to the Peltier element.
To achieve a wide range of contact line speeds (up to \SI{1}{\milli\meter\per\second}), we dispensed a series of consecutive drops at the same position. The pumping rates were systematically varied from \SI{6}{\micro\liter\per\second} to \SI{800}{\micro\liter\per\second} using a Legato 110 syringe pump (KD Scientific, Holliston, MA, USA).
Videos of the contact angle experiment were recorded with a Nikon D7500 (Minato, Japan) using a 60mm macro lens from Nikkor (Minato, Japan).
Contact angles were extracted from the videos using custom software \cite{Schubotz_Sessile_drop_analysis_Labbook_2025}.

\begin{figure}
	\centering
	\includegraphics[width=0.8\linewidth]{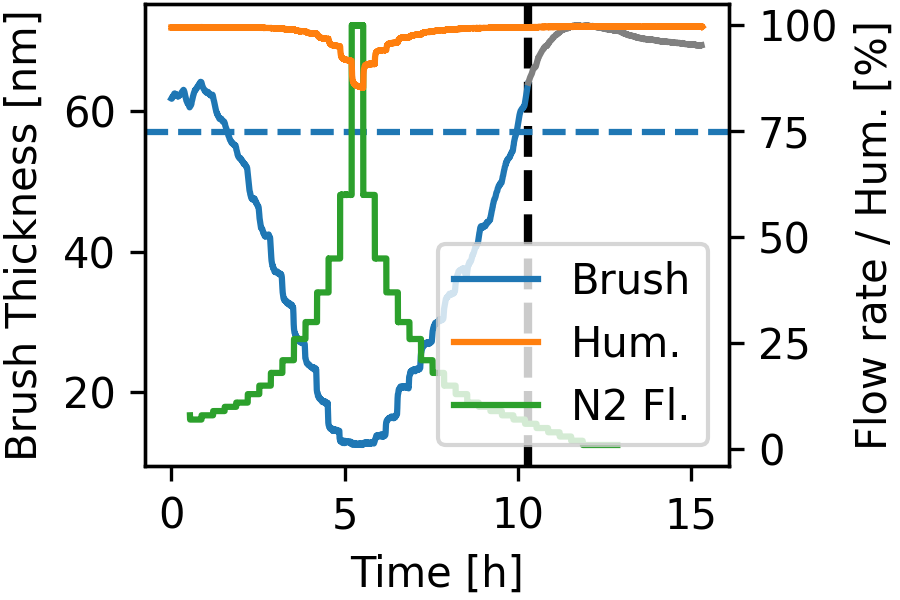}
	\caption{Typical experimental procedure for measuring the swelling isotherm at high RHs.
        The plot shows the measured brush thickness (blue line), the RH (orange line, right axis), and the applied dry N$_2$ flow rate (green line, right axis) as a function of time, where a 100\% flow rate corresponds to \SI{0,5}{L/min}.
        The experiment begins with a nearly saturated brush; modulating the N$_2$ flow rate induces deswelling and subsequent reswelling.
        The blue dashed line indicates the measured brush thickness in bulk liquid.
        The vertical black dotted line marks the onset of condensation, determined by the point where unphysical behavior appears in the optical model fit (see main text).}
	\label{fig:example_high_hum_start_0}
\end{figure}

The experiment begins by tuning the system to the point of full saturation (100\%~RH), balancing the in-cell vapor source with a modest influx of dry nitrogen ($\approx \SI{0.05}{\liter\per\minute}$). 
In the example shown in \autoref{fig:example_high_hum_start_0}, the initial thickness is $\approx$62~nm. 
This value is slightly greater than the brush's thickness in bulk liquid (57~nm) (blue dotted line), possibly due to a thin, optically unresolved layer of condensed water. 
This starting point ensures the brush is in a fully hydrated state at the start of the deswelling experiment ($t=0$).
By tuning the  nitrogen flow, we achieved a broad range of brush thicknesses. 
At a constant flow rate, the brush thickness remained stable for hours.
The nitrogen flow was adjusted every 20 minutes to allow the brush to reach equilibrium at each step.
Changes in flow rates were chosen to achieve approximately constant changes in brush thickness, resulting in small steps in flow rate at high degrees of swelling and large steps at low swelling, \autoref{fig:example_high_hum_start_0}.

After $ t \approx \SI{5}{\hour}$, we start reducing the nitrogen flow rate towards zero, which causes visible drops to condense on the brush surface at $ t \approx \SI{15}{\hour}$. 
To reliably identify the onset of condensation, we used the fitted refractive index of the brush, which decreases with increasing swelling. 
We identify the onset of condensation as the point in time at which this downward trend inverts and the fitted refractive index begins to increase due to incorrect fitting, occurring here at $t \approx \SI{10}{\hour}$. 
We neglected any data after this time. 
This criterion turned out to be in good agreement with a steep increase in the mean square error of the ellipsometer fit (see Fig. S4 in the SM~\cite{SM}). 

\begin{figure}
	\centering
	\includegraphics[width=0.8\linewidth]{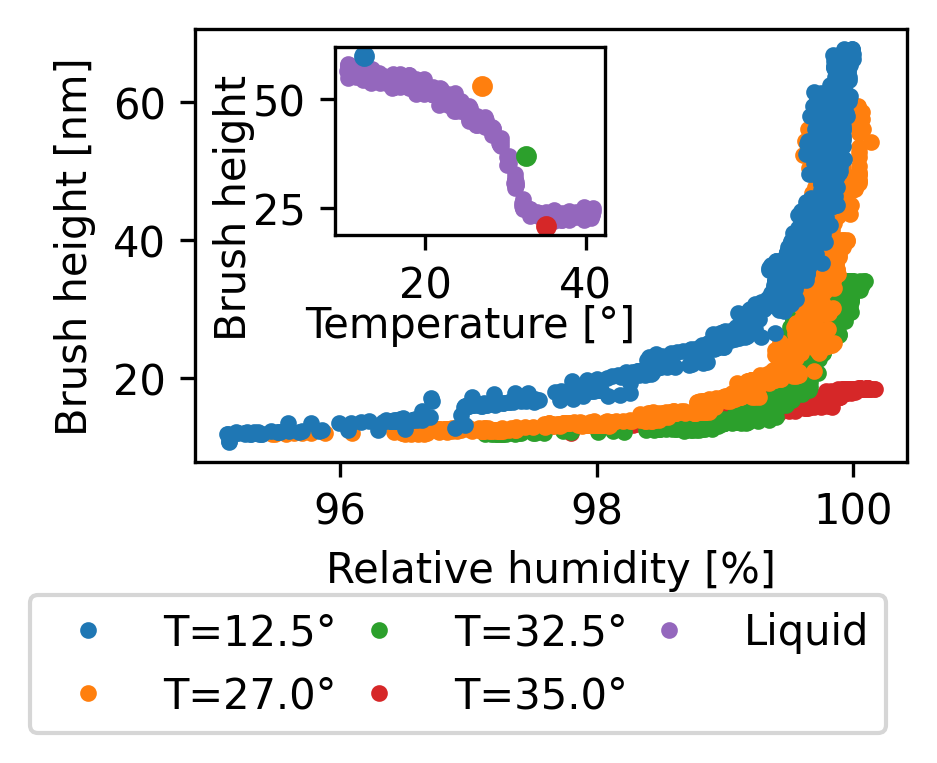}
	\caption{Swelling behavior for different humidities at different temperatures above and below the lower critical solution temperature.
            The inset compares swelling in the vapor phase to swelling in the liquid phase.}
	\label{fig:thickness_brush_gas_liq}
\end{figure}

The swelling of the polymer brush depends strongly on the temperature, \autoref{fig:thickness_brush_gas_liq}.
The LCST of PNiPAAm at \SI{\sim 32}{\degreeCelsius} resulted in greater swelling at temperatures below LCST and less swelling above LCST.
This temperature dependence is not limited to the final swollen state at 100\% RH; it also dictates the path to saturation. 
As shown in \autoref{fig:thickness_brush_gas_liq}, when approaching 100\% RH (e.g., 99\% RH), the brush is consistently more swollen at temperatures below the LCST. 

We used the same sample to perform temperature-dependent measurements of the brush thickness in liquid water, inset in \autoref{fig:thickness_brush_gas_liq}. 
Temperature-dependent swelling in water and saturated vapor are closely aligned, with slightly higher swelling in vapor. 
We attribute this minor difference to the stronger contrast in refractive index $n$ between the fully swollen brush ($n \approx 1.35$) and the gas phase ($n \approx 1.0$),  compared to the contrast between water ($n \approx 1.33$) and the brush.
Alternatively, a small systematic error in the precise experimental determination of the 100\%~RH point could also account for this difference.
Our experimental results clearly reject the Schr\"{o}der paradox for PNiPAAm. 

\begin{figure*}
	\centering
	\includegraphics[width=1\linewidth]{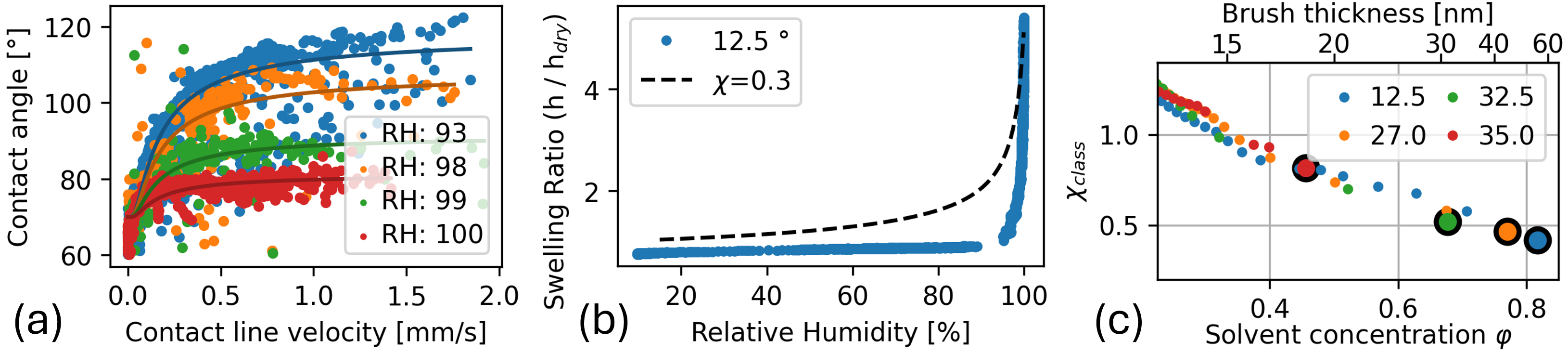}
	\caption{Experimental data, theoretical models, and the link to dynamic wetting.
    (a) The sharp swelling threshold is directly mirrored in the dynamic wetting, causing the advancing contact angle at high velocities to drop sharply only as ambient RH surpasses 98\%. The solid curves are fits using the adaptive wetting model~\cite{Butt2018}.
    (b) The experimental swelling isotherm at 12~$^\circ$C (blue dots) exhibits a sharp increase above 98\%~RH, a behavior not captured by the standard mean-field theory with any constant $\chi$ (e.g., black dashed line for $\chi=0.3$).
    (c) Our extended mean-field model explains this behavior via a positive feedback loop, revealed by the decrease in the derived classical interaction parameter, $\chi_\mathrm{class}$, with increasing solvent fraction $\varphi$. Large circles mark the state at 100\%~RH for each temperature.}
	\label{fig:Meanfield_modeling}
\end{figure*}

The sharp swelling of the brush above 98\%~RH directly dictates its dynamic wetting behavior. We demonstrate this with contact angle measurements of advancing water droplets at various contact line velocities, $U$, under different ambient RHs (\autoref{fig:Meanfield_modeling}a).

We find a distinct velocity-dependent crossover. At low velocities ($U < 0.1$~mm/s), the advancing contact angle is close to $\approx$70$^\circ$, independent of the ambient humidity. In this regime, the brush is always fully pre-swollen by a local, saturated vapor halo from the droplet itself.

At high velocities, the contact line outruns this local halo, and the ambient humidity becomes dominant. For low ambient RH, the advancing contact angle increases with velocity, attaining values as high as 110$^\circ$. In stark contrast, as the ambient RH increases above 98\%, the high-velocity contact angle drops dramatically. At 100\%~RH, the curve is nearly flat, rising only modestly to a maximum of $\approx$80$^\circ$.

This behavior is quantitatively described by the adaptive wetting model~\cite{Butt2018}, whose global fits to the data are shown as solid lines in \autoref{fig:Meanfield_modeling}a. In this framework, the analysis reveals that the surface energy parameter, $\Delta\gamma_\mathrm{SL}$, drops sharply for humidities above 98\% (see Fig. S5 in the SM~\cite{SM}), perfectly mirroring the sharp increase in the swelling isotherm. This quantitatively confirms the physical picture: pre-swelling the brush in a high-humidity environment changes its initial surface energy, thereby reducing the magnitude of the total interfacial energy change, $\Delta\gamma_\mathrm{SL}$, required for full adaptation of the brush to the drop. According to the adaptive wetting model, this smaller $\Delta\gamma_\mathrm{SL}$ directly results in a lower advancing contact angle at high velocities.

The sharp increase in swelling observed experimentally above \SI{98}{\percent}~RH (\autoref{fig:Meanfield_modeling}b) presents a significant challenge for standard theoretical models. The classic mean-field theory for polymer brushes~\cite{Lyatskaya94} finds the equilibrium swelling by balancing the chemical potential of the surrounding vapor, $\mu = k_\mathrm{B}T \ln(\mathrm{RH})$, with that derived from the brush's free energy. As demonstrated in \autoref{fig:Meanfield_modeling}b, this standard model, which assumes a constant Flory-Huggins interaction parameter, $\chi$, predicts a gradual swelling curve and is unable to reproduce the sharp increase we observe. This limitation is also confirmed by our coarse-grained molecular dynamics simulations (see Sec. S4 D in the SM~\cite{SM}). 

The systematic failure of constant-$\chi$ models motivates our extension of the theory. We assume that the interaction parameter is not constant but depends on the local solvent volume fraction within the brush, $\varphi$, making it a function $\chi(\varphi)$. This approach is inspired by previous work on the LCST of PNiPAAm~\cite{Afroze2000}. To analyze this, we reformulate the thermodynamics in terms of a more convenient potential: the grand potential free energy per monomer, $f(\varphi)$. This function is derived from the Helmholtz free energy via a Legendre transformation (see Sec. S4 A in the SM~\cite{SM} for details) and is given by:
\begin{equation}
    f(\varphi) = \frac{3}{2} \frac{\rho_g^2}{(1-\varphi)^2} + \frac{\varphi}{1-\varphi} \ln(\varphi) + \chi(\varphi) \cdot \varphi - \frac{\mu \cdot \varphi}{1-\varphi}
    \label{eq:free_energy_per_monomer}
\end{equation}
This equation can be separated into a term containing the unknown interaction parameter, $\eta(\varphi) = \chi(\varphi) \cdot \varphi$, and a known part, $f_0(\varphi, \mu)$, which contains all other terms. The equilibrium condition is found by minimizing this free energy with respect to the solvent concentration, $\partial f/\partial \varphi = 0$. This yields the differential equation:
\begin{equation}
    \frac{\partial f_0(\varphi, \mu)}{\partial \varphi} = -\chi(\varphi) - \varphi \frac{\mathrm{d}\chi(\varphi)}{\mathrm{d}\varphi}
    \label{eq:diff_eq}
\end{equation}
In contrast to previous approaches that assume a specific functional form for $\chi(\varphi)$ and fit its parameters, we solve this differential equation directly for $\chi(\varphi)$. As detailed in the SM~\cite{SM} (Sec. S4 B), the solution is:
\begin{equation}
    \chi(\varphi) = \frac{1}{\varphi} \left( f_0(\varphi=0, \mu) - f_0(\varphi, \mu) \right)
    \label{eq:chi_solution}
\end{equation}
\autoref{eq:chi_solution} allows us to use our experimental data to determine the interaction parameter $\chi(\varphi)$ without assuming a specific functional form. This calculation requires the solvent volume fraction, $\varphi$, which is derived from the measured brush thickness. To convert the measured thickness, $h$, to $\varphi$, we must assume a value for the dry brush thickness, $h_\mathrm{dry}$, as $\varphi = 1 - h_\mathrm{dry}/h$. Based on our measurements at very low RH ($\approx$\SI{10}{\percent}~RH), where the brush thickness is approximately \SI{11}{nm}, we assume a dry thickness of $h_\mathrm{dry} = \SI{10}{nm}$. While there is a small uncertainty in this value, its influence on the calculated $\chi$ diminishes rapidly as the brush swells to its final thickness of $\approx$\SI{60}{nm}.

The model also requires the dimensionless grafting density, $\rho_g$. We choose $\rho_g = 0.025$. This value is not a free fitting parameter but is a physically representative value for a polymer brush synthesized via the ``grafting-to'' method, and is consistent with the parameters used in our molecular dynamics simulations, which show a realistic chain conformation (see Sec. S4 F in the SM~\cite{SM} for a detailed justification).

It is important to distinguish our locally-defined interaction function, $\chi(\varphi)$, from the classical Flory-Huggins parameter, $\chi_\mathrm{class}$, which governs the overall solvent quality (e.g., the theta-point at $\chi_\mathrm{class}=0.5$). The relationship between them is found by comparing the equilibrium conditions for the two theoretical descriptions. For a model with a constant interaction parameter, the equilibrium condition is $\partial f_0/\partial\varphi = -\chi_\mathrm{class}$. For our model with a concentration-dependent parameter, the condition is given by \autoref{eq:diff_eq}. Equating these two expressions for $\partial f_0/\partial\varphi$ directly yields the formula for the effective classical parameter, a relationship previously explored by Baulin et al.~\cite{Baulin2003a}: $\chi_\mathrm{class}(\varphi) = \chi(\varphi) + \varphi (\mathrm{d}\chi/\mathrm{d}\varphi)$. To compute $\chi_\mathrm{class}$, which requires a stable derivative, the raw experimental data for each temperature were first binned into 200 intervals along the RH axis and then processed using a hybrid smoothing procedure (see Sec. S4 E in the SM~\cite{SM} for details). \autoref{fig:Meanfield_modeling}c illustrates the resulting $\chi_\mathrm{class}(\varphi)$ curves for four temperatures, with the points at full saturation highlighted. At 100\%~RH, the solvent quality improves at lower temperatures, consistent with the known LCST behavior of PNiPAAm. Specifically, for temperatures below the LCST ($\approx$\SI{32}{\celsius}), the calculated $\chi_\mathrm{class}$ values fall below the theta-point threshold of 0.5, indicating a good solvent state. Although the calculated $\chi_\mathrm{class}(\varphi)$ curves in \autoref{fig:Meanfield_modeling}c appear visually similar across different temperatures, the underlying $\chi(\varphi)$ functions from which they are derived are distinct. As discussed in the SM~\cite{SM} (Sec. S4 C), these subtle differences are crucial as they are the reason for the different swelling maxima observed at each temperature.

The shape of the $\chi_\mathrm{class}(\varphi)$ curves reveals the mechanism behind the sharp swelling. As the brush swells and the solvent fraction $\varphi$ increases, the value of $\chi$ decreases, indicating that the solvent quality improves. This creates a positive feedback loop: initial swelling improves the local polymer-solvent interaction, which in turn drives further, more rapid swelling. This positive feedback mechanism explains the sharp increase in brush thickness observed as the RH approaches \SI{100}{\percent}. This finding suggests a general principle for highly responsive systems: a strong, implicit dependence of the Flory-Huggins parameter on concentration can fundamentally alter the shape of the swelling isotherm, leading to sharp swelling of the brush.

In conclusion, we resolve the century-old Schröder paradox for PNiPAAm brushes, demonstrating that their equilibrium swelling in a saturated vapor matches the liquid state. Our key finding is that this occurs not gradually, but via a sharp increase in swelling concentrated above 98\% RH. This behavior, not captured by standard models, is explained by an extended mean-field theory that reveals the underlying mechanism: a positive feedback loop between swelling and solvent quality. 
This thermodynamic behavior is directly mirrored in the dynamic wetting, creating a velocity-dependent crossover where the advancing contact angle is governed by ambient humidity at high speeds, but solely by the droplet's local vapor halo at low speeds. Our work thus provides a unified picture of responsive interfaces, showing how a thermodynamic feedback mechanism dictates the equilibrium swelling state, which in turn governs the material's dynamic response to a moving contact line.

\begin{acknowledgments}
Simon Schubotz, Petra Uhlmann and Günter K. Auernhammer thank the Deutsche Forschungsgemeinschaft (DFG) for funding of project 422852551 (AU321/10-1,  UH121/3-1) within the priority program 2171, Quinn A. Besford thanks the DFG for funding project 496201730 (BE 7737/2-1) and Jens-Uwe Sommer thanks the DFG for funding project SO277/17-1. We thank Sebastian Rauch for synthesizing PNiPAAm for this project.
We acknowledge the fruitful discussions with Andreas Fery.
This text was written with the help of LLMs such as DeepL Write, Grammarly, ChatGPT, and Google Gemini.
\end{acknowledgments}

\bibliography{Articel}

\begin{thebibliography}{56}%
\makeatletter
\providecommand \@ifxundefined [1]{%
 \@ifx{#1\undefined}
}%
\providecommand \@ifnum [1]{%
 \ifnum #1\expandafter \@firstoftwo
 \else \expandafter \@secondoftwo
 \fi
}%
\providecommand \@ifx [1]{%
 \ifx #1\expandafter \@firstoftwo
 \else \expandafter \@secondoftwo
 \fi
}%
\providecommand \natexlab [1]{#1}%
\providecommand \enquote  [1]{``#1''}%
\providecommand \bibnamefont  [1]{#1}%
\providecommand \bibfnamefont [1]{#1}%
\providecommand \citenamefont [1]{#1}%
\providecommand \href@noop [0]{\@secondoftwo}%
\providecommand \href [0]{\begingroup \@sanitize@url \@href}%
\providecommand \@href[1]{\@@startlink{#1}\@@href}%
\providecommand \@@href[1]{\endgroup#1\@@endlink}%
\providecommand \@sanitize@url [0]{\catcode `\\12\catcode `\$12\catcode
  `\&12\catcode `\#12\catcode `\^12\catcode `\_12\catcode `\%12\relax}%
\providecommand \@@startlink[1]{}%
\providecommand \@@endlink[0]{}%
\providecommand \url  [0]{\begingroup\@sanitize@url \@url }%
\providecommand \@url [1]{\endgroup\@href {#1}{\urlprefix }}%
\providecommand \urlprefix  [0]{URL }%
\providecommand \Eprint [0]{\href }%
\providecommand \doibase [0]{https://doi.org/}%
\providecommand \selectlanguage [0]{\@gobble}%
\providecommand \bibinfo  [0]{\@secondoftwo}%
\providecommand \bibfield  [0]{\@secondoftwo}%
\providecommand \translation [1]{[#1]}%
\providecommand \BibitemOpen [0]{}%
\providecommand \bibitemStop [0]{}%
\providecommand \bibitemNoStop [0]{.\EOS\space}%
\providecommand \EOS [0]{\spacefactor3000\relax}%
\providecommand \BibitemShut  [1]{\csname bibitem#1\endcsname}%
\let\auto@bib@innerbib\@empty
\bibitem [{\citenamefont {Kuhn}\ \emph {et~al.}(1950)\citenamefont {Kuhn},
  \citenamefont {Hargitay}, \citenamefont {Katchalsky},\ and\ \citenamefont
  {Eisenberg}}]{W.1950}%
  \BibitemOpen
  \bibfield  {author} {\bibinfo {author} {\bibfnamefont {W.}~\bibnamefont
  {Kuhn}}, \bibinfo {author} {\bibfnamefont {B.}~\bibnamefont {Hargitay}},
  \bibinfo {author} {\bibfnamefont {A.}~\bibnamefont {Katchalsky}},\ and\
  \bibinfo {author} {\bibfnamefont {H.}~\bibnamefont {Eisenberg}},\ }\bibfield
  {title} {\bibinfo {title} {Reversible dilation and contraction by changing
  the state of ionization of high-polymer acid networks},\ }\href
  {https://doi.org/10.1038/165514a0} {\bibfield  {journal} {\bibinfo  {journal}
  {Nature}\ }\textbf {\bibinfo {volume} {165}},\ \bibinfo {pages} {514}
  (\bibinfo {year} {1950})}\BibitemShut {NoStop}%
\bibitem [{\citenamefont {Karg}\ \emph {et~al.}(2008)\citenamefont {Karg},
  \citenamefont {Pastoriza-Santos}, \citenamefont {Rodriguez-González},
  \citenamefont {von Klitzing}, \citenamefont {Wellert},\ and\ \citenamefont
  {Hellweg}}]{Karg2008}%
  \BibitemOpen
  \bibfield  {author} {\bibinfo {author} {\bibfnamefont {M.}~\bibnamefont
  {Karg}}, \bibinfo {author} {\bibfnamefont {I.}~\bibnamefont
  {Pastoriza-Santos}}, \bibinfo {author} {\bibfnamefont {B.}~\bibnamefont
  {Rodriguez-González}}, \bibinfo {author} {\bibfnamefont {R.}~\bibnamefont
  {von Klitzing}}, \bibinfo {author} {\bibfnamefont {S.}~\bibnamefont
  {Wellert}},\ and\ \bibinfo {author} {\bibfnamefont {T.}~\bibnamefont
  {Hellweg}},\ }\bibfield  {title} {\bibinfo {title} {Temperature, ph, and
  ionic strength induced changes of the swelling behavior of
  pnipam−poly(allylacetic acid) copolymer microgels},\ }\href
  {https://doi.org/10.1021/la702996p} {\bibfield  {journal} {\bibinfo
  {journal} {Langmuir}\ }\textbf {\bibinfo {volume} {24}},\ \bibinfo {pages}
  {6300} (\bibinfo {year} {2008})}\BibitemShut {NoStop}%
\bibitem [{\citenamefont {Taylor}\ and\ \citenamefont
  {Cerankowski}(1975)}]{Taylor1975}%
  \BibitemOpen
  \bibfield  {author} {\bibinfo {author} {\bibfnamefont {L.~D.}\ \bibnamefont
  {Taylor}}\ and\ \bibinfo {author} {\bibfnamefont {L.~D.}\ \bibnamefont
  {Cerankowski}},\ }\bibfield  {title} {\bibinfo {title} {Preparation of films
  exhibiting a balanced temperature dependence to permeation by aqueous
  solutions{\textemdash}a study of lower consolute behavior},\ }\href
  {https://doi.org/10.1002/pol.1975.170131113} {\bibfield  {journal} {\bibinfo
  {journal} {Journal of Polymer Science: Polymer Chemistry Edition}\ }\textbf
  {\bibinfo {volume} {13}},\ \bibinfo {pages} {2551} (\bibinfo {year}
  {1975})}\BibitemShut {NoStop}%
\bibitem [{\citenamefont {Akiyama}\ \emph {et~al.}(2004)\citenamefont
  {Akiyama}, \citenamefont {Kikuchi}, \citenamefont {Yamato},\ and\
  \citenamefont {Okano}}]{Akiyama2004}%
  \BibitemOpen
  \bibfield  {author} {\bibinfo {author} {\bibfnamefont {Y.}~\bibnamefont
  {Akiyama}}, \bibinfo {author} {\bibfnamefont {A.}~\bibnamefont {Kikuchi}},
  \bibinfo {author} {\bibfnamefont {M.}~\bibnamefont {Yamato}},\ and\ \bibinfo
  {author} {\bibfnamefont {T.}~\bibnamefont {Okano}},\ }\bibfield  {title}
  {\bibinfo {title} {Ultrathin poly(n-isopropylacrylamide) grafted layer on
  polystyrene surfaces for cell adhesion/detachment control},\ }\href
  {https://doi.org/10.1021/la036139f} {\bibfield  {journal} {\bibinfo
  {journal} {Langmuir}\ }\textbf {\bibinfo {volume} {20}},\ \bibinfo {pages}
  {5506} (\bibinfo {year} {2004})}\BibitemShut {NoStop}%
\bibitem [{\citenamefont {Plamper}\ and\ \citenamefont
  {Richtering}(2017)}]{Plamper2017}%
  \BibitemOpen
  \bibfield  {author} {\bibinfo {author} {\bibfnamefont {F.~A.}\ \bibnamefont
  {Plamper}}\ and\ \bibinfo {author} {\bibfnamefont {W.}~\bibnamefont
  {Richtering}},\ }\bibfield  {title} {\bibinfo {title} {Functional microgels
  and microgel systems},\ }\href {https://doi.org/10.1021/acs.accounts.6b00544}
  {\bibfield  {journal} {\bibinfo  {journal} {Accounts of Chemical Research}\
  }\textbf {\bibinfo {volume} {50}},\ \bibinfo {pages} {131} (\bibinfo {year}
  {2017})}\BibitemShut {NoStop}%
\bibitem [{\citenamefont {Kollarigowda}\ \emph {et~al.}(2017)\citenamefont
  {Kollarigowda}, \citenamefont {Fedele}, \citenamefont {Rianna}, \citenamefont
  {Calabuig}, \citenamefont {Manikas}, \citenamefont {Pagliarulo},
  \citenamefont {Ferraro}, \citenamefont {Cavalli},\ and\ \citenamefont
  {Netti}}]{Kollarigowda2017}%
  \BibitemOpen
  \bibfield  {author} {\bibinfo {author} {\bibfnamefont {R.~H.}\ \bibnamefont
  {Kollarigowda}}, \bibinfo {author} {\bibfnamefont {C.}~\bibnamefont
  {Fedele}}, \bibinfo {author} {\bibfnamefont {C.}~\bibnamefont {Rianna}},
  \bibinfo {author} {\bibfnamefont {A.}~\bibnamefont {Calabuig}}, \bibinfo
  {author} {\bibfnamefont {A.~C.}\ \bibnamefont {Manikas}}, \bibinfo {author}
  {\bibfnamefont {V.}~\bibnamefont {Pagliarulo}}, \bibinfo {author}
  {\bibfnamefont {P.}~\bibnamefont {Ferraro}}, \bibinfo {author} {\bibfnamefont
  {S.}~\bibnamefont {Cavalli}},\ and\ \bibinfo {author} {\bibfnamefont {P.~A.}\
  \bibnamefont {Netti}},\ }\bibfield  {title} {\bibinfo {title}
  {Light-responsive polymer brushes: active topographic cues for cell culture
  applications},\ }\href {https://doi.org/10.1039/c7py00462a} {\bibfield
  {journal} {\bibinfo  {journal} {Polymer Chemistry}\ }\textbf {\bibinfo
  {volume} {8}},\ \bibinfo {pages} {3271} (\bibinfo {year} {2017})}\BibitemShut
  {NoStop}%
\bibitem [{\citenamefont {Ishihara}\ \emph {et~al.}(1984)\citenamefont
  {Ishihara}, \citenamefont {Hamada}, \citenamefont {Kato},\ and\ \citenamefont
  {Shinohara}}]{Ishihara1984}%
  \BibitemOpen
  \bibfield  {author} {\bibinfo {author} {\bibfnamefont {K.}~\bibnamefont
  {Ishihara}}, \bibinfo {author} {\bibfnamefont {N.}~\bibnamefont {Hamada}},
  \bibinfo {author} {\bibfnamefont {S.}~\bibnamefont {Kato}},\ and\ \bibinfo
  {author} {\bibfnamefont {I.}~\bibnamefont {Shinohara}},\ }\bibfield  {title}
  {\bibinfo {title} {Photoresponse of the release behavior of an organic
  compound by a azoaromatic polymer device},\ }\href
  {https://doi.org/10.1002/pol.1984.170220341} {\bibfield  {journal} {\bibinfo
  {journal} {Journal of Polymer Science: Polymer Chemistry Edition}\ }\textbf
  {\bibinfo {volume} {22}},\ \bibinfo {pages} {881} (\bibinfo {year}
  {1984})}\BibitemShut {NoStop}%
\bibitem [{\citenamefont {Groten}\ \emph {et~al.}(2012)\citenamefont {Groten},
  \citenamefont {Bunte},\ and\ \citenamefont {Rühe}}]{Groten2012}%
  \BibitemOpen
  \bibfield  {author} {\bibinfo {author} {\bibfnamefont {J.}~\bibnamefont
  {Groten}}, \bibinfo {author} {\bibfnamefont {C.}~\bibnamefont {Bunte}},\ and\
  \bibinfo {author} {\bibfnamefont {J.}~\bibnamefont {Rühe}},\ }\bibfield
  {title} {\bibinfo {title} {Light-induced switching of surfaces at wetting
  transitions through photoisomerization of polymer monolayers},\ }\href
  {https://doi.org/10.1021/la302764k} {\bibfield  {journal} {\bibinfo
  {journal} {Langmuir}\ }\textbf {\bibinfo {volume} {28}},\ \bibinfo {pages}
  {15038} (\bibinfo {year} {2012})}\BibitemShut {NoStop}%
\bibitem [{\citenamefont {Heskins}\ and\ \citenamefont
  {Guillet}(1968)}]{Heskins1968}%
  \BibitemOpen
  \bibfield  {author} {\bibinfo {author} {\bibfnamefont {M.}~\bibnamefont
  {Heskins}}\ and\ \bibinfo {author} {\bibfnamefont {J.~E.}\ \bibnamefont
  {Guillet}},\ }\bibfield  {title} {\bibinfo {title} {Solution properties of
  poly(n-isopropylacrylamide)},\ }\href
  {https://doi.org/10.1080/10601326808051910} {\bibfield  {journal} {\bibinfo
  {journal} {Journal of Macromolecular Science: Part A - Chemistry}\ }\textbf
  {\bibinfo {volume} {2}},\ \bibinfo {pages} {1441} (\bibinfo {year}
  {1968})}\BibitemShut {NoStop}%
\bibitem [{\citenamefont {Stuart}\ \emph {et~al.}(2010)\citenamefont {Stuart},
  \citenamefont {Huck}, \citenamefont {Genzer}, \citenamefont {Müller},
  \citenamefont {Ober}, \citenamefont {Stamm}, \citenamefont {Sukhorukov},
  \citenamefont {Szleifer}, \citenamefont {Tsukruk}, \citenamefont {Urban},
  \citenamefont {Winnik}, \citenamefont {Zauscher}, \citenamefont {Luzinov},\
  and\ \citenamefont {Minko}}]{Stuart2010}%
  \BibitemOpen
  \bibfield  {author} {\bibinfo {author} {\bibfnamefont {M.~A.~C.}\
  \bibnamefont {Stuart}}, \bibinfo {author} {\bibfnamefont {W.~T.~S.}\
  \bibnamefont {Huck}}, \bibinfo {author} {\bibfnamefont {J.}~\bibnamefont
  {Genzer}}, \bibinfo {author} {\bibfnamefont {M.}~\bibnamefont {Müller}},
  \bibinfo {author} {\bibfnamefont {C.}~\bibnamefont {Ober}}, \bibinfo {author}
  {\bibfnamefont {M.}~\bibnamefont {Stamm}}, \bibinfo {author} {\bibfnamefont
  {G.~B.}\ \bibnamefont {Sukhorukov}}, \bibinfo {author} {\bibfnamefont
  {I.}~\bibnamefont {Szleifer}}, \bibinfo {author} {\bibfnamefont {V.~V.}\
  \bibnamefont {Tsukruk}}, \bibinfo {author} {\bibfnamefont {M.}~\bibnamefont
  {Urban}}, \bibinfo {author} {\bibfnamefont {F.}~\bibnamefont {Winnik}},
  \bibinfo {author} {\bibfnamefont {S.}~\bibnamefont {Zauscher}}, \bibinfo
  {author} {\bibfnamefont {I.}~\bibnamefont {Luzinov}},\ and\ \bibinfo {author}
  {\bibfnamefont {S.}~\bibnamefont {Minko}},\ }\bibfield  {title} {\bibinfo
  {title} {Emerging applications of stimuli-responsive polymer materials},\
  }\href {https://doi.org/10.1038/nmat2614} {\bibfield  {journal} {\bibinfo
  {journal} {Nature Materials}\ }\textbf {\bibinfo {volume} {9}},\ \bibinfo
  {pages} {101} (\bibinfo {year} {2010})}\BibitemShut {NoStop}%
\bibitem [{\citenamefont {GIL}\ and\ \citenamefont {HUDSON}(2004)}]{E2004}%
  \BibitemOpen
  \bibfield  {author} {\bibinfo {author} {\bibfnamefont {E.}~\bibnamefont
  {GIL}}\ and\ \bibinfo {author} {\bibfnamefont {S.}~\bibnamefont {HUDSON}},\
  }\bibfield  {title} {\bibinfo {title} {Stimuli-reponsive polymers and their
  bioconjugates},\ }\href {https://doi.org/10.1016/j.progpolymsci.2004.08.003}
  {\bibfield  {journal} {\bibinfo  {journal} {Progress in Polymer Science}\
  }\textbf {\bibinfo {volume} {29}},\ \bibinfo {pages} {1173} (\bibinfo {year}
  {2004})}\BibitemShut {NoStop}%
\bibitem [{\citenamefont {Magin}\ \emph {et~al.}(2010)\citenamefont {Magin},
  \citenamefont {Cooper},\ and\ \citenamefont {Brennan}}]{Magin2010}%
  \BibitemOpen
  \bibfield  {author} {\bibinfo {author} {\bibfnamefont {C.~M.}\ \bibnamefont
  {Magin}}, \bibinfo {author} {\bibfnamefont {S.~P.}\ \bibnamefont {Cooper}},\
  and\ \bibinfo {author} {\bibfnamefont {A.~B.}\ \bibnamefont {Brennan}},\
  }\bibfield  {title} {\bibinfo {title} {Non-toxic antifouling strategies},\
  }\href {https://doi.org/10.1016/s1369-7021(10)70058-4} {\bibfield  {journal}
  {\bibinfo  {journal} {Materials Today}\ }\textbf {\bibinfo {volume} {13}},\
  \bibinfo {pages} {36} (\bibinfo {year} {2010})}\BibitemShut {NoStop}%
\bibitem [{\citenamefont {Demirci}\ \emph {et~al.}(2017)\citenamefont
  {Demirci}, \citenamefont {Kinali-Demirci},\ and\ \citenamefont
  {Jiang}}]{Demirci:2017aa}%
  \BibitemOpen
  \bibfield  {author} {\bibinfo {author} {\bibfnamefont {S.}~\bibnamefont
  {Demirci}}, \bibinfo {author} {\bibfnamefont {S.}~\bibnamefont
  {Kinali-Demirci}},\ and\ \bibinfo {author} {\bibfnamefont {S.}~\bibnamefont
  {Jiang}},\ }\bibfield  {title} {\bibinfo {title} {A switchable polymer brush
  system for antifouling and controlled detection},\ }\href
  {https://doi.org/10.1039/c7cc00193b} {\bibfield  {journal} {\bibinfo
  {journal} {Chem Commun (Camb)}\ }\textbf {\bibinfo {volume} {53}},\ \bibinfo
  {pages} {3713} (\bibinfo {year} {2017})}\BibitemShut {NoStop}%
\bibitem [{\citenamefont {Belfort}\ \emph {et~al.}(1994)\citenamefont
  {Belfort}, \citenamefont {Davis},\ and\ \citenamefont
  {Zydney}}]{Belfort1994}%
  \BibitemOpen
  \bibfield  {author} {\bibinfo {author} {\bibfnamefont {G.}~\bibnamefont
  {Belfort}}, \bibinfo {author} {\bibfnamefont {R.~H.}\ \bibnamefont {Davis}},\
  and\ \bibinfo {author} {\bibfnamefont {A.~L.}\ \bibnamefont {Zydney}},\
  }\bibfield  {title} {\bibinfo {title} {The behavior of suspensions and
  macromolecular solutions in crossflow microfiltration},\ }\href
  {https://doi.org/10.1016/0376-7388(94)00119-7} {\bibfield  {journal}
  {\bibinfo  {journal} {Journal of Membrane Science}\ }\textbf {\bibinfo
  {volume} {96}},\ \bibinfo {pages} {1} (\bibinfo {year} {1994})}\BibitemShut
  {NoStop}%
\bibitem [{\citenamefont {Zhou}\ \emph {et~al.}(2018)\citenamefont {Zhou},
  \citenamefont {Zhao}, \citenamefont {Guo}, \citenamefont {Zhang},\ and\
  \citenamefont {Yu}}]{Zhou2018}%
  \BibitemOpen
  \bibfield  {author} {\bibinfo {author} {\bibfnamefont {X.}~\bibnamefont
  {Zhou}}, \bibinfo {author} {\bibfnamefont {F.}~\bibnamefont {Zhao}}, \bibinfo
  {author} {\bibfnamefont {Y.}~\bibnamefont {Guo}}, \bibinfo {author}
  {\bibfnamefont {Y.}~\bibnamefont {Zhang}},\ and\ \bibinfo {author}
  {\bibfnamefont {G.}~\bibnamefont {Yu}},\ }\bibfield  {title} {\bibinfo
  {title} {A hydrogel-based antifouling solar evaporator for highly efficient
  water desalination},\ }\href {https://doi.org/10.1039/c8ee00567b} {\bibfield
  {journal} {\bibinfo  {journal} {Energy \&amp; Environmental Science}\
  }\textbf {\bibinfo {volume} {11}},\ \bibinfo {pages} {1985} (\bibinfo {year}
  {2018})}\BibitemShut {NoStop}%
\bibitem [{\citenamefont {Qiu}\ and\ \citenamefont {Park}(2001)}]{Qiu2001}%
  \BibitemOpen
  \bibfield  {author} {\bibinfo {author} {\bibfnamefont {Y.}~\bibnamefont
  {Qiu}}\ and\ \bibinfo {author} {\bibfnamefont {K.}~\bibnamefont {Park}},\
  }\bibfield  {title} {\bibinfo {title} {Environment-sensitive hydrogels for
  drug delivery},\ }\href {https://doi.org/10.1016/s0169-409x(01)00203-4}
  {\bibfield  {journal} {\bibinfo  {journal} {Advanced Drug Delivery Reviews}\
  }\textbf {\bibinfo {volume} {53}},\ \bibinfo {pages} {321} (\bibinfo {year}
  {2001})}\BibitemShut {NoStop}%
\bibitem [{\citenamefont {Islam}\ \emph {et~al.}(2015)\citenamefont {Islam},
  \citenamefont {Xie}, \citenamefont {Huang}, \citenamefont {Smyth},\ and\
  \citenamefont {Serpe}}]{Islam2015}%
  \BibitemOpen
  \bibfield  {author} {\bibinfo {author} {\bibfnamefont {M.~R.}\ \bibnamefont
  {Islam}}, \bibinfo {author} {\bibfnamefont {S.}~\bibnamefont {Xie}}, \bibinfo
  {author} {\bibfnamefont {D.}~\bibnamefont {Huang}}, \bibinfo {author}
  {\bibfnamefont {K.}~\bibnamefont {Smyth}},\ and\ \bibinfo {author}
  {\bibfnamefont {M.~J.}\ \bibnamefont {Serpe}},\ }\bibfield  {title} {\bibinfo
  {title} {Poly (n-isopropylacrylamide) microgel-based optical devices for
  humidity sensing},\ }\href {https://doi.org/10.1016/j.aca.2015.09.039}
  {\bibfield  {journal} {\bibinfo  {journal} {Analytica Chimica Acta}\ }\textbf
  {\bibinfo {volume} {898}},\ \bibinfo {pages} {101} (\bibinfo {year}
  {2015})}\BibitemShut {NoStop}%
\bibitem [{\citenamefont {Schönberg}\ \emph {et~al.}(2016)\citenamefont
  {Schönberg}, \citenamefont {Kondrashov}, \citenamefont {Prokhorov},\ and\
  \citenamefont {Rühe}}]{Schoenberg2016}%
  \BibitemOpen
  \bibfield  {author} {\bibinfo {author} {\bibfnamefont {J.-N.}\ \bibnamefont
  {Schönberg}}, \bibinfo {author} {\bibfnamefont {V.}~\bibnamefont
  {Kondrashov}}, \bibinfo {author} {\bibfnamefont {A.}~\bibnamefont
  {Prokhorov}},\ and\ \bibinfo {author} {\bibfnamefont {J.}~\bibnamefont
  {Rühe}},\ }\bibfield  {title} {\bibinfo {title} {Capacitive humidity and
  dew-point sensing: Influence of wetting of surface-attached polymer
  monolayers on the sensor response},\ }\href
  {https://doi.org/10.1016/j.snb.2015.08.021} {\bibfield  {journal} {\bibinfo
  {journal} {Sensors and Actuators B: Chemical}\ }\textbf {\bibinfo {volume}
  {222}},\ \bibinfo {pages} {87} (\bibinfo {year} {2016})}\BibitemShut
  {NoStop}%
\bibitem [{\citenamefont {Chen}\ \emph {et~al.}(2010)\citenamefont {Chen},
  \citenamefont {Ferris}, \citenamefont {Zhang}, \citenamefont {Ducker},\ and\
  \citenamefont {Zauscher}}]{Chen:2010af}%
  \BibitemOpen
  \bibfield  {author} {\bibinfo {author} {\bibfnamefont {T.}~\bibnamefont
  {Chen}}, \bibinfo {author} {\bibfnamefont {R.}~\bibnamefont {Ferris}},
  \bibinfo {author} {\bibfnamefont {J.}~\bibnamefont {Zhang}}, \bibinfo
  {author} {\bibfnamefont {R.}~\bibnamefont {Ducker}},\ and\ \bibinfo {author}
  {\bibfnamefont {S.}~\bibnamefont {Zauscher}},\ }\bibfield  {title} {\bibinfo
  {title} {Stimulus-responsive polymer brushes on surfaces: Transduction
  mechanisms and applications},\ }\href
  {https://doi.org/10.1016/j.progpolymsci.2009.11.004} {\bibfield  {journal}
  {\bibinfo  {journal} {Progress in Polymer Science}\ }\textbf {\bibinfo
  {volume} {35}},\ \bibinfo {pages} {94} (\bibinfo {year} {2010})}\BibitemShut
  {NoStop}%
\bibitem [{\citenamefont {Poppinga}\ \emph {et~al.}(2017)\citenamefont
  {Poppinga}, \citenamefont {Zollfrank}, \citenamefont {Prucker}, \citenamefont
  {Rühe}, \citenamefont {Menges}, \citenamefont {Cheng},\ and\ \citenamefont
  {Speck}}]{Poppinga2017}%
  \BibitemOpen
  \bibfield  {author} {\bibinfo {author} {\bibfnamefont {S.}~\bibnamefont
  {Poppinga}}, \bibinfo {author} {\bibfnamefont {C.}~\bibnamefont {Zollfrank}},
  \bibinfo {author} {\bibfnamefont {O.}~\bibnamefont {Prucker}}, \bibinfo
  {author} {\bibfnamefont {J.}~\bibnamefont {Rühe}}, \bibinfo {author}
  {\bibfnamefont {A.}~\bibnamefont {Menges}}, \bibinfo {author} {\bibfnamefont
  {T.}~\bibnamefont {Cheng}},\ and\ \bibinfo {author} {\bibfnamefont
  {T.}~\bibnamefont {Speck}},\ }\bibfield  {title} {\bibinfo {title} {Toward a
  new generation of smart biomimetic actuators for architecture},\ }\bibfield
  {journal} {\bibinfo  {journal} {Advanced Materials}\ }\textbf {\bibinfo
  {volume} {30}},\ \href {https://doi.org/10.1002/adma.201703653}
  {10.1002/adma.201703653} (\bibinfo {year} {2017})\BibitemShut {NoStop}%
\bibitem [{\citenamefont {Galaev}(1999)}]{Galaev1999}%
  \BibitemOpen
  \bibfield  {author} {\bibinfo {author} {\bibfnamefont {I.}~\bibnamefont
  {Galaev}},\ }\bibfield  {title} {\bibinfo {title} {“smart” polymers and
  what they could do in biotechnology and medicine},\ }\href
  {https://doi.org/10.1016/s0167-7799(99)01345-1} {\bibfield  {journal}
  {\bibinfo  {journal} {Trends in Biotechnology}\ }\textbf {\bibinfo {volume}
  {17}},\ \bibinfo {pages} {335} (\bibinfo {year} {1999})}\BibitemShut
  {NoStop}%
\bibitem [{\citenamefont {de~Gennes}(1980)}]{Gennes1980}%
  \BibitemOpen
  \bibfield  {author} {\bibinfo {author} {\bibfnamefont {P.~G.}\ \bibnamefont
  {de~Gennes}},\ }\bibfield  {title} {\bibinfo {title} {Conformations of
  polymers attached to an interface},\ }\href
  {https://doi.org/10.1021/ma60077a009} {\bibfield  {journal} {\bibinfo
  {journal} {Macromolecules}\ }\textbf {\bibinfo {volume} {13}},\ \bibinfo
  {pages} {1069} (\bibinfo {year} {1980})}\BibitemShut {NoStop}%
\bibitem [{\citenamefont {MILNER}(1991)}]{MILNER1991}%
  \BibitemOpen
  \bibfield  {author} {\bibinfo {author} {\bibfnamefont {S.~T.}\ \bibnamefont
  {MILNER}},\ }\bibfield  {title} {\bibinfo {title} {Polymer brushes},\ }\href
  {https://doi.org/10.1126/science.251.4996.905} {\bibfield  {journal}
  {\bibinfo  {journal} {Science}\ }\textbf {\bibinfo {volume} {251}},\ \bibinfo
  {pages} {905} (\bibinfo {year} {1991})}\BibitemShut {NoStop}%
\bibitem [{\citenamefont {Brown}\ \emph {et~al.}(1990)\citenamefont {Brown},
  \citenamefont {Char},\ and\ \citenamefont {Deline}}]{Brown_1990aa}%
  \BibitemOpen
  \bibfield  {author} {\bibinfo {author} {\bibfnamefont {H.~R.}\ \bibnamefont
  {Brown}}, \bibinfo {author} {\bibfnamefont {K.}~\bibnamefont {Char}},\ and\
  \bibinfo {author} {\bibfnamefont {V.~R.}\ \bibnamefont {Deline}},\ }\bibfield
   {title} {\bibinfo {title} {Enthalpy-driven swelling of a polymer brush},\
  }\href@noop {} {\bibfield  {journal} {\bibinfo  {journal} {Macromolecules}\
  }\textbf {\bibinfo {volume} {23}},\ \bibinfo {pages} {3383} (\bibinfo {year}
  {1990})}\BibitemShut {NoStop}%
\bibitem [{\citenamefont {Wittmer}\ \emph {et~al.}(1994)\citenamefont
  {Wittmer}, \citenamefont {Johner},\ and\ \citenamefont
  {Joanny}}]{WITTMER199485}%
  \BibitemOpen
  \bibfield  {author} {\bibinfo {author} {\bibfnamefont {J.}~\bibnamefont
  {Wittmer}}, \bibinfo {author} {\bibfnamefont {A.}~\bibnamefont {Johner}},\
  and\ \bibinfo {author} {\bibfnamefont {J.}~\bibnamefont {Joanny}},\
  }\bibfield  {title} {\bibinfo {title} {Some dynamic properties of grafted
  polymer layers},\ }\href@noop {} {\bibfield  {journal} {\bibinfo  {journal}
  {Colloids and Surfaces A: Physicochemical and Engineering Aspects}\ }\textbf
  {\bibinfo {volume} {86}},\ \bibinfo {pages} {85} (\bibinfo {year}
  {1994})}\BibitemShut {NoStop}%
\bibitem [{\citenamefont {Tsujii}\ \emph {et~al.}(2006)\citenamefont {Tsujii},
  \citenamefont {Ohno}, \citenamefont {Yamamoto}, \citenamefont {Goto},\ and\
  \citenamefont {Fukuda}}]{Tsujii2006}%
  \BibitemOpen
  \bibfield  {author} {\bibinfo {author} {\bibfnamefont {Y.}~\bibnamefont
  {Tsujii}}, \bibinfo {author} {\bibfnamefont {K.}~\bibnamefont {Ohno}},
  \bibinfo {author} {\bibfnamefont {S.}~\bibnamefont {Yamamoto}}, \bibinfo
  {author} {\bibfnamefont {A.}~\bibnamefont {Goto}},\ and\ \bibinfo {author}
  {\bibfnamefont {T.}~\bibnamefont {Fukuda}},\ }\bibinfo {title} {Structure and
  properties of high-density polymer brushes prepared by
  surface-initiatedliving radical polymerization}\ (\bibinfo  {publisher}
  {Springer Berlin Heidelberg},\ \bibinfo {address} {Berlin, Heidelberg},\
  \bibinfo {year} {2006})\ pp.\ \bibinfo {pages} {1--45}\BibitemShut {NoStop}%
\bibitem [{\citenamefont {Advincula}\ \emph {et~al.}(2004)\citenamefont
  {Advincula}, \citenamefont {Brittain}, \citenamefont {Caster},\ and\
  \citenamefont {R{\"u}he}}]{2004a}%
  \BibitemOpen
  \bibinfo {editor} {\bibfnamefont {R.~C.}\ \bibnamefont {Advincula}}, \bibinfo
  {editor} {\bibfnamefont {W.~J.}\ \bibnamefont {Brittain}}, \bibinfo {editor}
  {\bibfnamefont {K.~C.}\ \bibnamefont {Caster}},\ and\ \bibinfo {editor}
  {\bibfnamefont {J.}~\bibnamefont {R{\"u}he}},\ eds.,\ \href
  {https://doi.org/10.1002/3527603824} {\emph {\bibinfo {title} {Polymer
  Brushes: Synthesis, Characterization, Applications}}}\ (\bibinfo  {publisher}
  {Wiley},\ \bibinfo {year} {2004})\BibitemShut {NoStop}%
\bibitem [{\citenamefont {Birshtein}\ and\ \citenamefont
  {Lyatskaya}(1994{\natexlab{a}})}]{Birshtein1994}%
  \BibitemOpen
  \bibfield  {author} {\bibinfo {author} {\bibfnamefont {T.~M.}\ \bibnamefont
  {Birshtein}}\ and\ \bibinfo {author} {\bibfnamefont {Y.~V.}\ \bibnamefont
  {Lyatskaya}},\ }\bibfield  {title} {\bibinfo {title} {Theory of the
  collapse-stretching transition of a polymer brush in a mixed solvent},\
  }\href {https://doi.org/10.1021/ma00083a028} {\bibfield  {journal} {\bibinfo
  {journal} {Macromolecules}\ }\textbf {\bibinfo {volume} {27}},\ \bibinfo
  {pages} {1256} (\bibinfo {year} {1994}{\natexlab{a}})}\BibitemShut {NoStop}%
\bibitem [{\citenamefont {Ritsema~van Eck}\ \emph
  {et~al.}(2022{\natexlab{a}})\citenamefont {Ritsema~van Eck}, \citenamefont
  {Chiappisi},\ and\ \citenamefont {de~Beer}}]{Eck2022}%
  \BibitemOpen
  \bibfield  {author} {\bibinfo {author} {\bibfnamefont {G.~C.}\ \bibnamefont
  {Ritsema~van Eck}}, \bibinfo {author} {\bibfnamefont {L.}~\bibnamefont
  {Chiappisi}},\ and\ \bibinfo {author} {\bibfnamefont {S.}~\bibnamefont
  {de~Beer}},\ }\bibfield  {title} {\bibinfo {title} {Fundamentals and
  applications of polymer brushes in air},\ }\href
  {https://doi.org/10.1021/acsapm.1c01615} {\bibfield  {journal} {\bibinfo
  {journal} {ACS Applied Polymer Materials}\ }\textbf {\bibinfo {volume} {4}},\
  \bibinfo {pages} {3062} (\bibinfo {year} {2022}{\natexlab{a}})},\ \Eprint
  {https://arxiv.org/abs/https://doi.org/10.1021/acsapm.1c01615}
  {https://doi.org/10.1021/acsapm.1c01615} \BibitemShut {NoStop}%
\bibitem [{\citenamefont {Chen}\ \emph {et~al.}(2023)\citenamefont {Chen},
  \citenamefont {Chen},\ and\ \citenamefont {Tao}}]{CHEN2023113050}%
  \BibitemOpen
  \bibfield  {author} {\bibinfo {author} {\bibfnamefont {L.}~\bibnamefont
  {Chen}}, \bibinfo {author} {\bibfnamefont {Y.}~\bibnamefont {Chen}},\ and\
  \bibinfo {author} {\bibfnamefont {W.-Q.}\ \bibnamefont {Tao}},\ }\bibfield
  {title} {\bibinfo {title} {Schroeder's paradox in proton exchange membrane
  fuel cells: A review},\ }\href@noop {} {\bibfield  {journal} {\bibinfo
  {journal} {Renewable and Sustainable Energy Reviews}\ }\textbf {\bibinfo
  {volume} {173}},\ \bibinfo {pages} {113050} (\bibinfo {year}
  {2023})}\BibitemShut {NoStop}%
\bibitem [{\citenamefont {Onishi}\ \emph {et~al.}(2007)\citenamefont {Onishi},
  \citenamefont {Prausnitz},\ and\ \citenamefont {Newman}}]{Onishi_2007aa}%
  \BibitemOpen
  \bibfield  {author} {\bibinfo {author} {\bibfnamefont {L.~M.}\ \bibnamefont
  {Onishi}}, \bibinfo {author} {\bibfnamefont {J.~M.}\ \bibnamefont
  {Prausnitz}},\ and\ \bibinfo {author} {\bibfnamefont {J.}~\bibnamefont
  {Newman}},\ }\bibfield  {title} {\bibinfo {title} {Water−nafion equilibria.
  absence of schroeder's paradox},\ }\href@noop {} {\bibfield  {journal}
  {\bibinfo  {journal} {The Journal of Physical Chemistry B}\ }\textbf
  {\bibinfo {volume} {111}},\ \bibinfo {pages} {10166} (\bibinfo {year}
  {2007})}\BibitemShut {NoStop}%
\bibitem [{\citenamefont {Beers}\ \emph {et~al.}(2014)\citenamefont {Beers},
  \citenamefont {Yakovlev}, \citenamefont {Jackson}, \citenamefont {Wang},
  \citenamefont {Hexemer}, \citenamefont {Downing},\ and\ \citenamefont
  {Balsara}}]{Beers_2014aa}%
  \BibitemOpen
  \bibfield  {author} {\bibinfo {author} {\bibfnamefont {K.~M.}\ \bibnamefont
  {Beers}}, \bibinfo {author} {\bibfnamefont {S.}~\bibnamefont {Yakovlev}},
  \bibinfo {author} {\bibfnamefont {A.}~\bibnamefont {Jackson}}, \bibinfo
  {author} {\bibfnamefont {X.}~\bibnamefont {Wang}}, \bibinfo {author}
  {\bibfnamefont {A.}~\bibnamefont {Hexemer}}, \bibinfo {author} {\bibfnamefont
  {K.~H.}\ \bibnamefont {Downing}},\ and\ \bibinfo {author} {\bibfnamefont
  {N.~P.}\ \bibnamefont {Balsara}},\ }\bibfield  {title} {\bibinfo {title}
  {Absence of schroeder's paradox in a nanostructured block copolymer
  electrolyte membrane},\ }\href@noop {} {\bibfield  {journal} {\bibinfo
  {journal} {The Journal of Physical Chemistry B}\ }\textbf {\bibinfo {volume}
  {118}},\ \bibinfo {pages} {6785} (\bibinfo {year} {2014})}\BibitemShut
  {NoStop}%
\bibitem [{\citenamefont {Jeck}\ \emph {et~al.}(2011)\citenamefont {Jeck},
  \citenamefont {Scharfer},\ and\ \citenamefont {Kind}}]{JECK201174}%
  \BibitemOpen
  \bibfield  {author} {\bibinfo {author} {\bibfnamefont {S.}~\bibnamefont
  {Jeck}}, \bibinfo {author} {\bibfnamefont {P.}~\bibnamefont {Scharfer}},\
  and\ \bibinfo {author} {\bibfnamefont {M.}~\bibnamefont {Kind}},\ }\bibfield
  {title} {\bibinfo {title} {Absence of schroeder's paradox: Experimental
  evidence for water-swollen nafion{\textregistered} membranes},\ }\href@noop
  {} {\bibfield  {journal} {\bibinfo  {journal} {Journal of Membrane Science}\
  }\textbf {\bibinfo {volume} {373}},\ \bibinfo {pages} {74} (\bibinfo {year}
  {2011})}\BibitemShut {NoStop}%
\bibitem [{\citenamefont {Vallieres}\ \emph {et~al.}(2006)\citenamefont
  {Vallieres}, \citenamefont {Winkelmann}, \citenamefont {Roizard},
  \citenamefont {Favre}, \citenamefont {Scharfer},\ and\ \citenamefont
  {Kind}}]{VALLIERES2006357}%
  \BibitemOpen
  \bibfield  {author} {\bibinfo {author} {\bibfnamefont {C.}~\bibnamefont
  {Vallieres}}, \bibinfo {author} {\bibfnamefont {D.}~\bibnamefont
  {Winkelmann}}, \bibinfo {author} {\bibfnamefont {D.}~\bibnamefont {Roizard}},
  \bibinfo {author} {\bibfnamefont {E.}~\bibnamefont {Favre}}, \bibinfo
  {author} {\bibfnamefont {P.}~\bibnamefont {Scharfer}},\ and\ \bibinfo
  {author} {\bibfnamefont {M.}~\bibnamefont {Kind}},\ }\bibfield  {title}
  {\bibinfo {title} {On schroeder's paradox},\ }\href@noop {} {\bibfield
  {journal} {\bibinfo  {journal} {Journal of Membrane Science}\ }\textbf
  {\bibinfo {volume} {278}},\ \bibinfo {pages} {357} (\bibinfo {year}
  {2006})}\BibitemShut {NoStop}%
\bibitem [{\citenamefont {Eikerling}\ and\ \citenamefont
  {Berg}(2011)}]{C1SM05273J}%
  \BibitemOpen
  \bibfield  {author} {\bibinfo {author} {\bibfnamefont {M.~H.}\ \bibnamefont
  {Eikerling}}\ and\ \bibinfo {author} {\bibfnamefont {P.}~\bibnamefont
  {Berg}},\ }\bibfield  {title} {\bibinfo {title} {Poroelectroelastic theory of
  water sorption and swelling in polymer electrolyte membranes},\ }\href@noop
  {} {\bibfield  {journal} {\bibinfo  {journal} {Soft Matter}\ }\textbf
  {\bibinfo {volume} {7}},\ \bibinfo {pages} {5976} (\bibinfo {year}
  {2011})}\BibitemShut {NoStop}%
\bibitem [{\citenamefont {Hartmann}\ \emph {et~al.}(2024)\citenamefont
  {Hartmann}, \citenamefont {Diekmann}, \citenamefont {Greve},\ and\
  \citenamefont {Thiele}}]{Hartmann_2024aa}%
  \BibitemOpen
  \bibfield  {author} {\bibinfo {author} {\bibfnamefont {S.}~\bibnamefont
  {Hartmann}}, \bibinfo {author} {\bibfnamefont {J.}~\bibnamefont {Diekmann}},
  \bibinfo {author} {\bibfnamefont {D.}~\bibnamefont {Greve}},\ and\ \bibinfo
  {author} {\bibfnamefont {U.}~\bibnamefont {Thiele}},\ }\bibfield  {title}
  {\bibinfo {title} {Drops on polymer brushes: Advances in thin-film modeling
  of adaptive substrates},\ }\href@noop {} {\bibfield  {journal} {\bibinfo
  {journal} {Langmuir}\ }\textbf {\bibinfo {volume} {40}},\ \bibinfo {pages}
  {4001} (\bibinfo {year} {2024})}\BibitemShut {NoStop}%
\bibitem [{\citenamefont {Kubota}\ \emph {et~al.}(1990)\citenamefont {Kubota},
  \citenamefont {Fujishige},\ and\ \citenamefont {Ando}}]{Kubota1990}%
  \BibitemOpen
  \bibfield  {author} {\bibinfo {author} {\bibfnamefont {K.}~\bibnamefont
  {Kubota}}, \bibinfo {author} {\bibfnamefont {S.}~\bibnamefont {Fujishige}},\
  and\ \bibinfo {author} {\bibfnamefont {I.}~\bibnamefont {Ando}},\ }\bibfield
  {title} {\bibinfo {title} {Solution properties of poly(n-isopropylacrylamide)
  in water},\ }\href {https://doi.org/10.1295/polymj.22.15} {\bibfield
  {journal} {\bibinfo  {journal} {Polymer Journal}\ }\textbf {\bibinfo {volume}
  {22}},\ \bibinfo {pages} {15} (\bibinfo {year} {1990})}\BibitemShut {NoStop}%
\bibitem [{\citenamefont {Bittrich}\ \emph {et~al.}(2012)\citenamefont
  {Bittrich}, \citenamefont {Burkert}, \citenamefont {Mueller}, \citenamefont
  {Eichhorn}, \citenamefont {Stamm},\ and\ \citenamefont
  {Uhlmann}}]{Bittrich2012}%
  \BibitemOpen
  \bibfield  {author} {\bibinfo {author} {\bibfnamefont {E.}~\bibnamefont
  {Bittrich}}, \bibinfo {author} {\bibfnamefont {S.}~\bibnamefont {Burkert}},
  \bibinfo {author} {\bibfnamefont {M.}~\bibnamefont {Mueller}}, \bibinfo
  {author} {\bibfnamefont {K.-J.}\ \bibnamefont {Eichhorn}}, \bibinfo {author}
  {\bibfnamefont {M.}~\bibnamefont {Stamm}},\ and\ \bibinfo {author}
  {\bibfnamefont {P.}~\bibnamefont {Uhlmann}},\ }\bibfield  {title} {\bibinfo
  {title} {Temperature-sensitive swelling of poly (n-isopropylacrylamide)
  brushes with low molecular weight and grafting density},\ }\href
  {https://doi.org/10.1021/la204230a} {\bibfield  {journal} {\bibinfo
  {journal} {Langmuir}\ }\textbf {\bibinfo {volume} {28}},\ \bibinfo {pages}
  {3439} (\bibinfo {year} {2012})}\BibitemShut {NoStop}%
\bibitem [{\citenamefont {Afroze}\ \emph {et~al.}(2000)\citenamefont {Afroze},
  \citenamefont {Nies},\ and\ \citenamefont {Berghmans}}]{Afroze2000}%
  \BibitemOpen
  \bibfield  {author} {\bibinfo {author} {\bibfnamefont {F.}~\bibnamefont
  {Afroze}}, \bibinfo {author} {\bibfnamefont {E.}~\bibnamefont {Nies}},\ and\
  \bibinfo {author} {\bibfnamefont {H.}~\bibnamefont {Berghmans}},\ }\bibfield
  {title} {\bibinfo {title} {Phase transitions in the system
  poly(n-isopropylacrylamide)/water and swelling behaviour of the corresponding
  networks},\ }\href {https://doi.org/10.1016/s0022-2860(00)00559-7} {\bibfield
   {journal} {\bibinfo  {journal} {Journal of Molecular Structure}\ }\textbf
  {\bibinfo {volume} {554}},\ \bibinfo {pages} {55} (\bibinfo {year}
  {2000})}\BibitemShut {NoStop}%
\bibitem [{\citenamefont {Halperin}\ \emph {et~al.}(2015)\citenamefont
  {Halperin}, \citenamefont {Kröger},\ and\ \citenamefont
  {Winnik}}]{Halperin2015}%
  \BibitemOpen
  \bibfield  {author} {\bibinfo {author} {\bibfnamefont {A.}~\bibnamefont
  {Halperin}}, \bibinfo {author} {\bibfnamefont {M.}~\bibnamefont {Kröger}},\
  and\ \bibinfo {author} {\bibfnamefont {F.~M.}\ \bibnamefont {Winnik}},\
  }\bibfield  {title} {\bibinfo {title} {Poly(n‐isopropylacrylamide) phase
  diagrams: Fifty years of research},\ }\href
  {https://doi.org/10.1002/anie.201506663} {\bibfield  {journal} {\bibinfo
  {journal} {Angewandte Chemie International Edition}\ }\textbf {\bibinfo
  {volume} {54}},\ \bibinfo {pages} {15342} (\bibinfo {year}
  {2015})}\BibitemShut {NoStop}%
\bibitem [{\citenamefont {Butt}\ \emph {et~al.}(2018)\citenamefont {Butt},
  \citenamefont {Berger}, \citenamefont {Steffen}, \citenamefont {Vollmer},\
  and\ \citenamefont {Weber}}]{Butt2018}%
  \BibitemOpen
  \bibfield  {author} {\bibinfo {author} {\bibfnamefont {H.-J.}\ \bibnamefont
  {Butt}}, \bibinfo {author} {\bibfnamefont {R.}~\bibnamefont {Berger}},
  \bibinfo {author} {\bibfnamefont {W.}~\bibnamefont {Steffen}}, \bibinfo
  {author} {\bibfnamefont {D.}~\bibnamefont {Vollmer}},\ and\ \bibinfo {author}
  {\bibfnamefont {S.~A.~L.}\ \bibnamefont {Weber}},\ }\bibfield  {title}
  {\bibinfo {title} {Adaptive wetting{\textemdash}adaptation in wetting},\
  }\href {https://doi.org/10.1021/acs.langmuir.8b01783} {\bibfield  {journal}
  {\bibinfo  {journal} {Langmuir}\ }\textbf {\bibinfo {volume} {34}},\ \bibinfo
  {pages} {11292} (\bibinfo {year} {2018})}\BibitemShut {NoStop}%
\bibitem [{\citenamefont {Schubotz}\ \emph {et~al.}(2021)\citenamefont
  {Schubotz}, \citenamefont {Honnigfort}, \citenamefont {Nazari}, \citenamefont
  {Fery}, \citenamefont {Sommer}, \citenamefont {Uhlmann}, \citenamefont
  {Braunschweig},\ and\ \citenamefont {Auernhammer}}]{Schubotz2021}%
  \BibitemOpen
  \bibfield  {author} {\bibinfo {author} {\bibfnamefont {S.}~\bibnamefont
  {Schubotz}}, \bibinfo {author} {\bibfnamefont {C.}~\bibnamefont
  {Honnigfort}}, \bibinfo {author} {\bibfnamefont {S.}~\bibnamefont {Nazari}},
  \bibinfo {author} {\bibfnamefont {A.}~\bibnamefont {Fery}}, \bibinfo {author}
  {\bibfnamefont {J.-U.}\ \bibnamefont {Sommer}}, \bibinfo {author}
  {\bibfnamefont {P.}~\bibnamefont {Uhlmann}}, \bibinfo {author} {\bibfnamefont
  {B.}~\bibnamefont {Braunschweig}},\ and\ \bibinfo {author} {\bibfnamefont
  {G.~K.}\ \bibnamefont {Auernhammer}},\ }\bibfield  {title} {\bibinfo {title}
  {Memory effects in polymer brushes showing co-nonsolvency effects},\ }\href
  {https://doi.org/10.1016/j.cis.2021.102442} {\bibfield  {journal} {\bibinfo
  {journal} {Advances in Colloid and Interface Science}\ }\textbf {\bibinfo
  {volume} {294}},\ \bibinfo {pages} {102442} (\bibinfo {year}
  {2021})}\BibitemShut {NoStop}%
\bibitem [{\citenamefont {Sychev}\ \emph {et~al.}(2023)\citenamefont {Sychev},
  \citenamefont {Schubotz}, \citenamefont {Besford}, \citenamefont {Fery},\
  and\ \citenamefont {Auernhammer}}]{Sychev2023}%
  \BibitemOpen
  \bibfield  {author} {\bibinfo {author} {\bibfnamefont {D.}~\bibnamefont
  {Sychev}}, \bibinfo {author} {\bibfnamefont {S.}~\bibnamefont {Schubotz}},
  \bibinfo {author} {\bibfnamefont {Q.~A.}\ \bibnamefont {Besford}}, \bibinfo
  {author} {\bibfnamefont {A.}~\bibnamefont {Fery}},\ and\ \bibinfo {author}
  {\bibfnamefont {G.~K.}\ \bibnamefont {Auernhammer}},\ }\bibfield  {title}
  {\bibinfo {title} {Critical analysis of adhesion work measurements from
  afm-based techniques for soft contact},\ }\href
  {https://doi.org/10.1016/j.jcis.2023.03.139} {\bibfield  {journal} {\bibinfo
  {journal} {Journal of Colloid and Interface Science}\ }\textbf {\bibinfo
  {volume} {642}},\ \bibinfo {pages} {216} (\bibinfo {year}
  {2023})}\BibitemShut {NoStop}%
\bibitem [{\citenamefont {Schubotz}\ \emph {et~al.}(2023)\citenamefont
  {Schubotz}, \citenamefont {Besford}, \citenamefont {Nazari}, \citenamefont
  {Uhlmann}, \citenamefont {Bittrich}, \citenamefont {Sommer},\ and\
  \citenamefont {Auernhammer}}]{Schubotz2023}%
  \BibitemOpen
  \bibfield  {author} {\bibinfo {author} {\bibfnamefont {S.}~\bibnamefont
  {Schubotz}}, \bibinfo {author} {\bibfnamefont {Q.~A.}\ \bibnamefont
  {Besford}}, \bibinfo {author} {\bibfnamefont {S.}~\bibnamefont {Nazari}},
  \bibinfo {author} {\bibfnamefont {P.}~\bibnamefont {Uhlmann}}, \bibinfo
  {author} {\bibfnamefont {E.}~\bibnamefont {Bittrich}}, \bibinfo {author}
  {\bibfnamefont {J.-U.}\ \bibnamefont {Sommer}},\ and\ \bibinfo {author}
  {\bibfnamefont {G.~K.}\ \bibnamefont {Auernhammer}},\ }\bibfield  {title}
  {\bibinfo {title} {Influence of the atmosphere on the wettability of polymer
  brushes},\ }\href {https://doi.org/10.1021/acs.langmuir.2c03009} {\bibfield
  {journal} {\bibinfo  {journal} {Langmuir}\ }\textbf {\bibinfo {volume}
  {39}},\ \bibinfo {pages} {4872} (\bibinfo {year} {2023})}\BibitemShut
  {NoStop}%
\bibitem [{\citenamefont {Diekmann}\ and\ \citenamefont
  {Thiele}(2024)}]{Diekmann2024}%
  \BibitemOpen
  \bibfield  {author} {\bibinfo {author} {\bibfnamefont {J.}~\bibnamefont
  {Diekmann}}\ and\ \bibinfo {author} {\bibfnamefont {U.}~\bibnamefont
  {Thiele}},\ }\bibfield  {title} {\bibinfo {title} {Drops of volatile binary
  mixtures on brush-covered substrates},\ }\href
  {https://doi.org/10.1140/epjs/s11734-024-01169-4} {\bibfield  {journal}
  {\bibinfo  {journal} {The European Physical Journal Special Topics}\ }\textbf
  {\bibinfo {volume} {233}},\ \bibinfo {pages} {1615} (\bibinfo {year}
  {2024})}\BibitemShut {NoStop}%
\bibitem [{\citenamefont {Kap}\ \emph {et~al.}(2023)\citenamefont {Kap},
  \citenamefont {Hartmann}, \citenamefont {Hoek}, \citenamefont {de~Beer},
  \citenamefont {Siretanu}, \citenamefont {Thiele},\ and\ \citenamefont
  {Mugele}}]{Kap_2023aa}%
  \BibitemOpen
  \bibfield  {author} {\bibinfo {author} {\bibfnamefont {{\"O}.}~\bibnamefont
  {Kap}}, \bibinfo {author} {\bibfnamefont {S.}~\bibnamefont {Hartmann}},
  \bibinfo {author} {\bibfnamefont {H.}~\bibnamefont {Hoek}}, \bibinfo {author}
  {\bibfnamefont {S.}~\bibnamefont {de~Beer}}, \bibinfo {author} {\bibfnamefont
  {I.}~\bibnamefont {Siretanu}}, \bibinfo {author} {\bibfnamefont
  {U.}~\bibnamefont {Thiele}},\ and\ \bibinfo {author} {\bibfnamefont
  {F.}~\bibnamefont {Mugele}},\ }\bibfield  {title} {\bibinfo {title}
  {{Nonequilibrium configurations of swelling polymer brush layers induced by
  spreading drops of weakly volatile oil}},\ }\href@noop {} {\bibfield
  {journal} {\bibinfo  {journal} {The Journal of Chemical Physics}\ }\textbf
  {\bibinfo {volume} {158}} (\bibinfo {year} {2023})}\BibitemShut {NoStop}%
\bibitem [{\citenamefont {Besford}\ \emph
  {et~al.}(2022{\natexlab{a}})\citenamefont {Besford}, \citenamefont {Merlitz},
  \citenamefont {Schubotz}, \citenamefont {Yong}, \citenamefont {Chae},
  \citenamefont {Schnepf}, \citenamefont {Weiss}, \citenamefont {Auernhammer},
  \citenamefont {Sommer}, \citenamefont {Uhlmann},\ and\ \citenamefont
  {Fery}}]{Besford2022}%
  \BibitemOpen
  \bibfield  {author} {\bibinfo {author} {\bibfnamefont {Q.~A.}\ \bibnamefont
  {Besford}}, \bibinfo {author} {\bibfnamefont {H.}~\bibnamefont {Merlitz}},
  \bibinfo {author} {\bibfnamefont {S.}~\bibnamefont {Schubotz}}, \bibinfo
  {author} {\bibfnamefont {H.}~\bibnamefont {Yong}}, \bibinfo {author}
  {\bibfnamefont {S.}~\bibnamefont {Chae}}, \bibinfo {author} {\bibfnamefont
  {M.~J.}\ \bibnamefont {Schnepf}}, \bibinfo {author} {\bibfnamefont
  {A.~C.~G.}\ \bibnamefont {Weiss}}, \bibinfo {author} {\bibfnamefont {G.~K.}\
  \bibnamefont {Auernhammer}}, \bibinfo {author} {\bibfnamefont {J.-U.}\
  \bibnamefont {Sommer}}, \bibinfo {author} {\bibfnamefont {P.}~\bibnamefont
  {Uhlmann}},\ and\ \bibinfo {author} {\bibfnamefont {A.}~\bibnamefont
  {Fery}},\ }\bibfield  {title} {\bibinfo {title} {Mechanofluorescent polymer
  brush surfaces that spatially resolve surface solvation},\ }\href
  {https://doi.org/10.1021/acsnano.2c00277} {\bibfield  {journal} {\bibinfo
  {journal} {ACS Nano}\ }\textbf {\bibinfo {volume} {16}},\ \bibinfo {pages}
  {3383} (\bibinfo {year} {2022}{\natexlab{a}})},\ \bibinfo {note} {pMID:
  35112848},\ \Eprint
  {https://arxiv.org/abs/https://doi.org/10.1021/acsnano.2c00277}
  {https://doi.org/10.1021/acsnano.2c00277} \BibitemShut {NoStop}%
\bibitem [{\citenamefont {Besford}\ \emph
  {et~al.}(2022{\natexlab{b}})\citenamefont {Besford}, \citenamefont
  {Schubotz}, \citenamefont {Chae}, \citenamefont {Özdabak Sert},
  \citenamefont {Weiss}, \citenamefont {Auernhammer}, \citenamefont {Uhlmann},
  \citenamefont {Farinha},\ and\ \citenamefont {Fery}}]{Besford2022a}%
  \BibitemOpen
  \bibfield  {author} {\bibinfo {author} {\bibfnamefont {Q.~A.}\ \bibnamefont
  {Besford}}, \bibinfo {author} {\bibfnamefont {S.}~\bibnamefont {Schubotz}},
  \bibinfo {author} {\bibfnamefont {S.}~\bibnamefont {Chae}}, \bibinfo {author}
  {\bibfnamefont {A.~B.}\ \bibnamefont {Özdabak Sert}}, \bibinfo {author}
  {\bibfnamefont {A.~C.~G.}\ \bibnamefont {Weiss}}, \bibinfo {author}
  {\bibfnamefont {G.~K.}\ \bibnamefont {Auernhammer}}, \bibinfo {author}
  {\bibfnamefont {P.}~\bibnamefont {Uhlmann}}, \bibinfo {author} {\bibfnamefont
  {J.~P.~S.}\ \bibnamefont {Farinha}},\ and\ \bibinfo {author} {\bibfnamefont
  {A.}~\bibnamefont {Fery}},\ }\bibfield  {title} {\bibinfo {title} {Molecular
  transport within polymer brushes: A {FRET} view at aqueous interfaces},\
  }\href {https://doi.org/10.3390/molecules27093043} {\bibfield  {journal}
  {\bibinfo  {journal} {Molecules}\ }\textbf {\bibinfo {volume} {27}},\
  \bibinfo {pages} {3043} (\bibinfo {year} {2022}{\natexlab{b}})}\BibitemShut
  {NoStop}%
\bibitem [{\citenamefont {Biesalski}\ and\ \citenamefont
  {R{\"u}he}(2000)}]{Biesalski:2000aa}%
  \BibitemOpen
  \bibfield  {author} {\bibinfo {author} {\bibfnamefont {M.}~\bibnamefont
  {Biesalski}}\ and\ \bibinfo {author} {\bibfnamefont {J.}~\bibnamefont
  {R{\"u}he}},\ }\bibfield  {title} {\bibinfo {title} {Swelling of a
  polyelectrolyte brush in humid air},\ }\href
  {https://doi.org/10.1021/la990863+} {\bibfield  {journal} {\bibinfo
  {journal} {Langmuir}\ }\textbf {\bibinfo {volume} {16}},\ \bibinfo {pages}
  {1943 } (\bibinfo {year} {2000})}\BibitemShut {NoStop}%
\bibitem [{\citenamefont {Ritsema~van Eck}\ \emph
  {et~al.}(2022{\natexlab{b}})\citenamefont {Ritsema~van Eck}, \citenamefont
  {Kiens}, \citenamefont {Veldscholte}, \citenamefont {Brió~Pérez},\ and\
  \citenamefont {de~Beer}}]{RitsemavanEck2022}%
  \BibitemOpen
  \bibfield  {author} {\bibinfo {author} {\bibfnamefont {G.~C.}\ \bibnamefont
  {Ritsema~van Eck}}, \bibinfo {author} {\bibfnamefont {E.~M.}\ \bibnamefont
  {Kiens}}, \bibinfo {author} {\bibfnamefont {L.~B.}\ \bibnamefont
  {Veldscholte}}, \bibinfo {author} {\bibfnamefont {M.}~\bibnamefont
  {Brió~Pérez}},\ and\ \bibinfo {author} {\bibfnamefont {S.}~\bibnamefont
  {de~Beer}},\ }\bibfield  {title} {\bibinfo {title} {Vapor swelling of polymer
  brushes compared to nongrafted films},\ }\href
  {https://doi.org/10.1021/acs.langmuir.2c01889} {\bibfield  {journal}
  {\bibinfo  {journal} {Langmuir}\ }\textbf {\bibinfo {volume} {38}},\ \bibinfo
  {pages} {13763} (\bibinfo {year} {2022}{\natexlab{b}})}\BibitemShut {NoStop}%
\bibitem [{SM()}]{SM}%
  \BibitemOpen
  \href@noop {} {}\bibinfo {note} {See Supplemental Material at [URL will be
  inserted by publisher] for full experimental details, data processing, and
  theoretical derivations.}\BibitemShut {Stop}%
\bibitem [{\citenamefont {Schubotz}\ \emph {et~al.}(2025)\citenamefont
  {Schubotz}, \citenamefont {Schubotz},\ and\ \citenamefont
  {Auernhammer}}]{schubotz2025}%
  \BibitemOpen
  \bibfield  {author} {\bibinfo {author} {\bibfnamefont {S.}~\bibnamefont
  {Schubotz}}, \bibinfo {author} {\bibfnamefont {M.}~\bibnamefont {Schubotz}},\
  and\ \bibinfo {author} {\bibfnamefont {G.~K.}\ \bibnamefont {Auernhammer}},\
  }\bibfield  {title} {\bibinfo {title} {Electronic {{Laboratory Notebook}}:
  {{An Adaptable Solution}}},\ }\href {https://doi.org/10.5334/jors.391}
  {\bibfield  {journal} {\bibinfo  {journal} {Journal of Open Research
  Software}\ }\textbf {\bibinfo {volume} {13}},\ \bibinfo {pages} {11}
  (\bibinfo {year} {2025})}\BibitemShut {NoStop}%
\bibitem [{\citenamefont {Alduchov}\ and\ \citenamefont
  {Eskridge}(1996)}]{Alduchov1996}%
  \BibitemOpen
  \bibfield  {author} {\bibinfo {author} {\bibfnamefont {O.~A.}\ \bibnamefont
  {Alduchov}}\ and\ \bibinfo {author} {\bibfnamefont {R.~E.}\ \bibnamefont
  {Eskridge}},\ }\bibfield  {title} {\bibinfo {title} {Improved magnus form
  approximation of saturation vapor pressure},\ }\href
  {http://www.jstor.org/stable/26187406} {\bibfield  {journal} {\bibinfo
  {journal} {Journal of Applied Meteorology (1988-2005)}\ }\textbf {\bibinfo
  {volume} {35}},\ \bibinfo {pages} {601} (\bibinfo {year} {1996})}\BibitemShut
  {NoStop}%
\bibitem [{\citenamefont
  {Schubotz}(2025)}]{Schubotz_Sessile_drop_analysis_Labbook_2025}%
  \BibitemOpen
  \bibfield  {author} {\bibinfo {author} {\bibfnamefont {S.}~\bibnamefont
  {Schubotz}},\ }\href {https://doi.org/10.5281/zenodo.16279486} {\bibinfo
  {title} {{Sessile.drop.analysis.Labbook}}} (\bibinfo {year} {2025}),\
  \bibinfo {note} {version 1.0.0. Also available at
  \url{https://github.com/SimonSchubotz/Sessile.drop.analysis.Labbook}}\BibitemShut
  {NoStop}%
\bibitem [{\citenamefont {Birshtein}\ and\ \citenamefont
  {Lyatskaya}(1994{\natexlab{b}})}]{Lyatskaya94}%
  \BibitemOpen
  \bibfield  {author} {\bibinfo {author} {\bibfnamefont {T.~M.}\ \bibnamefont
  {Birshtein}}\ and\ \bibinfo {author} {\bibfnamefont {Y.~V.}\ \bibnamefont
  {Lyatskaya}},\ }\bibfield  {title} {\bibinfo {title} {Theory of the
  collapse-stretching transition of a polymer brush in a mixed solvent},\
  }\href {https://doi.org/10.1021/ma00083a028} {\bibfield  {journal} {\bibinfo
  {journal} {Macromolecules}\ }\textbf {\bibinfo {volume} {27}},\ \bibinfo
  {pages} {1256} (\bibinfo {year} {1994}{\natexlab{b}})},\ \Eprint
  {https://arxiv.org/abs/https://doi.org/10.1021/ma00083a028}
  {https://doi.org/10.1021/ma00083a028} \BibitemShut {NoStop}%
\bibitem [{\citenamefont {Baulin}\ and\ \citenamefont
  {Halperin}(2003)}]{Baulin2003a}%
  \BibitemOpen
  \bibfield  {author} {\bibinfo {author} {\bibfnamefont {V.~A.}\ \bibnamefont
  {Baulin}}\ and\ \bibinfo {author} {\bibfnamefont {A.}~\bibnamefont
  {Halperin}},\ }\bibfield  {title} {\bibinfo {title} {Signatures of a
  concentration‐dependent flory χ parameter: Swelling and collapse of coils
  and brushes},\ }\href {https://doi.org/10.1002/mats.200350014} {\bibfield
  {journal} {\bibinfo  {journal} {Macromolecular Theory and Simulations}\
  }\textbf {\bibinfo {volume} {12}},\ \bibinfo {pages} {549} (\bibinfo {year}
  {2003})}\BibitemShut {NoStop}%
\end{thebibliography}%


\begin{thebibliography}{14}%
\makeatletter
\providecommand \@ifxundefined [1]{%
 \@ifx{#1\undefined}
}%
\providecommand \@ifnum [1]{%
 \ifnum #1\expandafter \@firstoftwo
 \else \expandafter \@secondoftwo
 \fi
}%
\providecommand \@ifx [1]{%
 \ifx #1\expandafter \@firstoftwo
 \else \expandafter \@secondoftwo
 \fi
}%
\providecommand \natexlab [1]{#1}%
\providecommand \enquote  [1]{``#1''}%
\providecommand \bibnamefont  [1]{#1}%
\providecommand \bibfnamefont [1]{#1}%
\providecommand \citenamefont [1]{#1}%
\providecommand \href@noop [0]{\@secondoftwo}%
\providecommand \href [0]{\begingroup \@sanitize@url \@href}%
\providecommand \@href[1]{\@@startlink{#1}\@@href}%
\providecommand \@@href[1]{\endgroup#1\@@endlink}%
\providecommand \@sanitize@url [0]{\catcode `\\12\catcode `\$12\catcode
  `\&12\catcode `\#12\catcode `\^12\catcode `\_12\catcode `\%12\relax}%
\providecommand \@@startlink[1]{}%
\providecommand \@@endlink[0]{}%
\providecommand \url  [0]{\begingroup\@sanitize@url \@url }%
\providecommand \@url [1]{\endgroup\@href {#1}{\urlprefix }}%
\providecommand \urlprefix  [0]{URL }%
\providecommand \Eprint [0]{\href }%
\providecommand \doibase [0]{https://doi.org/}%
\providecommand \selectlanguage [0]{\@gobble}%
\providecommand \bibinfo  [0]{\@secondoftwo}%
\providecommand \bibfield  [0]{\@secondoftwo}%
\providecommand \translation [1]{[#1]}%
\providecommand \BibitemOpen [0]{}%
\providecommand \bibitemStop [0]{}%
\providecommand \bibitemNoStop [0]{.\EOS\space}%
\providecommand \EOS [0]{\spacefactor3000\relax}%
\providecommand \BibitemShut  [1]{\csname bibitem#1\endcsname}%
\let\auto@bib@innerbib\@empty
\bibitem [{\citenamefont {Schubotz}\ \emph {et~al.}(2025)\citenamefont
  {Schubotz}, \citenamefont {Schubotz},\ and\ \citenamefont
  {Auernhammer}}]{schubotz2025}%
  \BibitemOpen
  \bibfield  {author} {\bibinfo {author} {\bibfnamefont {S.}~\bibnamefont
  {Schubotz}}, \bibinfo {author} {\bibfnamefont {M.}~\bibnamefont {Schubotz}},\
  and\ \bibinfo {author} {\bibfnamefont {G.~K.}\ \bibnamefont {Auernhammer}},\
  }\bibfield  {title} {\bibinfo {title} {Electronic {{Laboratory Notebook}}:
  {{An Adaptable Solution}}},\ }\href {https://doi.org/10.5334/jors.391}
  {\bibfield  {journal} {\bibinfo  {journal} {Journal of Open Research
  Software}\ }\textbf {\bibinfo {volume} {13}},\ \bibinfo {pages} {11}
  (\bibinfo {year} {2025})}\BibitemShut {NoStop}%
\bibitem [{\citenamefont {Schubotz}(2022)}]{SchubotzCode2022}%
  \BibitemOpen
  \bibfield  {author} {\bibinfo {author} {\bibfnamefont {S.}~\bibnamefont
  {Schubotz}},\ }\href@noop {} {\bibinfo {title}
  {Electronic-laboratory-notebook}} (\bibinfo {year} {2022}),\ \bibinfo {note}
  {available at
  \url{https://github.com/gipplab/Electronic-Laboratory-Notebook}, Accessed on
  16 September 2025}\BibitemShut {NoStop}%
\bibitem [{\citenamefont {Rauch}\ \emph {et~al.}(2012)\citenamefont {Rauch},
  \citenamefont {Eichhorn}, \citenamefont {Oertel}, \citenamefont {Stamm},
  \citenamefont {Kuckling},\ and\ \citenamefont {Uhlmann}}]{Rauch2012}%
  \BibitemOpen
  \bibfield  {author} {\bibinfo {author} {\bibfnamefont {S.}~\bibnamefont
  {Rauch}}, \bibinfo {author} {\bibfnamefont {K.-J.}\ \bibnamefont {Eichhorn}},
  \bibinfo {author} {\bibfnamefont {U.}~\bibnamefont {Oertel}}, \bibinfo
  {author} {\bibfnamefont {M.}~\bibnamefont {Stamm}}, \bibinfo {author}
  {\bibfnamefont {D.}~\bibnamefont {Kuckling}},\ and\ \bibinfo {author}
  {\bibfnamefont {P.}~\bibnamefont {Uhlmann}},\ }\bibfield  {title} {\bibinfo
  {title} {Temperature responsive polymer brushes with clicked rhodamine b:
  synthesis, characterization and swelling dynamics studied by spectroscopic
  ellipsometry},\ }\href {https://doi.org/10.1039/c2sm26571k} {\bibfield
  {journal} {\bibinfo  {journal} {Soft Matter}\ }\textbf {\bibinfo {volume}
  {8}},\ \bibinfo {pages} {10260} (\bibinfo {year} {2012})}\BibitemShut
  {NoStop}%
\bibitem [{\citenamefont {Zdyrko}\ \emph {et~al.}(2006)\citenamefont {Zdyrko},
  \citenamefont {Iyer},\ and\ \citenamefont {Luzinov}}]{Zdyrko2006}%
  \BibitemOpen
  \bibfield  {author} {\bibinfo {author} {\bibfnamefont {B.}~\bibnamefont
  {Zdyrko}}, \bibinfo {author} {\bibfnamefont {K.~S.}\ \bibnamefont {Iyer}},\
  and\ \bibinfo {author} {\bibfnamefont {I.}~\bibnamefont {Luzinov}},\
  }\bibfield  {title} {\bibinfo {title} {Macromolecular anchoring layers for
  polymer grafting: Comparative study},\ }\href
  {https://doi.org/10.1016/j.polymer.2005.11.029} {\bibfield  {journal}
  {\bibinfo  {journal} {Polymer}\ }\textbf {\bibinfo {volume} {47}},\ \bibinfo
  {pages} {272} (\bibinfo {year} {2006})}\BibitemShut {NoStop}%
\bibitem [{\citenamefont {Iyer}\ \emph {et~al.}(2003)\citenamefont {Iyer},
  \citenamefont {Zdyrko}, \citenamefont {Malz}, \citenamefont {Pionteck},\ and\
  \citenamefont {Luzinov}}]{Iyer2003}%
  \BibitemOpen
  \bibfield  {author} {\bibinfo {author} {\bibfnamefont {K.~S.}\ \bibnamefont
  {Iyer}}, \bibinfo {author} {\bibfnamefont {B.}~\bibnamefont {Zdyrko}},
  \bibinfo {author} {\bibfnamefont {H.}~\bibnamefont {Malz}}, \bibinfo {author}
  {\bibfnamefont {J.}~\bibnamefont {Pionteck}},\ and\ \bibinfo {author}
  {\bibfnamefont {I.}~\bibnamefont {Luzinov}},\ }\bibfield  {title} {\bibinfo
  {title} {Polystyrene layers grafted to macromolecular anchoring layer},\
  }\href {https://doi.org/10.1021/ma034460z} {\bibfield  {journal} {\bibinfo
  {journal} {Macromolecules}\ }\textbf {\bibinfo {volume} {36}},\ \bibinfo
  {pages} {6519} (\bibinfo {year} {2003})}\BibitemShut {NoStop}%
\bibitem [{\citenamefont {Bittrich}\ \emph {et~al.}(2012)\citenamefont
  {Bittrich}, \citenamefont {Burkert}, \citenamefont {Mueller}, \citenamefont
  {Eichhorn}, \citenamefont {Stamm},\ and\ \citenamefont
  {Uhlmann}}]{Bittrich2012}%
  \BibitemOpen
  \bibfield  {author} {\bibinfo {author} {\bibfnamefont {E.}~\bibnamefont
  {Bittrich}}, \bibinfo {author} {\bibfnamefont {S.}~\bibnamefont {Burkert}},
  \bibinfo {author} {\bibfnamefont {M.}~\bibnamefont {Mueller}}, \bibinfo
  {author} {\bibfnamefont {K.-J.}\ \bibnamefont {Eichhorn}}, \bibinfo {author}
  {\bibfnamefont {M.}~\bibnamefont {Stamm}},\ and\ \bibinfo {author}
  {\bibfnamefont {P.}~\bibnamefont {Uhlmann}},\ }\bibfield  {title} {\bibinfo
  {title} {Temperature-sensitive swelling of poly (n-isopropylacrylamide)
  brushes with low molecular weight and grafting density},\ }\href
  {https://doi.org/10.1021/la204230a} {\bibfield  {journal} {\bibinfo
  {journal} {Langmuir}\ }\textbf {\bibinfo {volume} {28}},\ \bibinfo {pages}
  {3439} (\bibinfo {year} {2012})}\BibitemShut {NoStop}%
\bibitem [{\citenamefont {Xue}\ \emph {et~al.}(2011)\citenamefont {Xue},
  \citenamefont {Yonet-Tanyeri}, \citenamefont {Brouette}, \citenamefont
  {Sferrazza}, \citenamefont {Braun},\ and\ \citenamefont
  {Leckband}}]{Xue2011}%
  \BibitemOpen
  \bibfield  {author} {\bibinfo {author} {\bibfnamefont {C.}~\bibnamefont
  {Xue}}, \bibinfo {author} {\bibfnamefont {N.}~\bibnamefont {Yonet-Tanyeri}},
  \bibinfo {author} {\bibfnamefont {N.}~\bibnamefont {Brouette}}, \bibinfo
  {author} {\bibfnamefont {M.}~\bibnamefont {Sferrazza}}, \bibinfo {author}
  {\bibfnamefont {P.~V.}\ \bibnamefont {Braun}},\ and\ \bibinfo {author}
  {\bibfnamefont {D.~E.}\ \bibnamefont {Leckband}},\ }\bibfield  {title}
  {\bibinfo {title} {Protein adsorption on poly(n-isopropylacrylamide) brushes:
  {{Dependence}} on grafting density and chain collapse},\ }\href
  {https://doi.org/10.1021/la2001909} {\bibfield  {journal} {\bibinfo
  {journal} {Langmuir}\ }\textbf {\bibinfo {volume} {27}},\ \bibinfo {pages}
  {8810} (\bibinfo {year} {2011})}\BibitemShut {NoStop}%
\bibitem [{\citenamefont {Schubotz}\ \emph {et~al.}(2023)\citenamefont
  {Schubotz}, \citenamefont {Besford}, \citenamefont {Nazari}, \citenamefont
  {Uhlmann}, \citenamefont {Bittrich}, \citenamefont {Sommer},\ and\
  \citenamefont {Auernhammer}}]{Schubotz2023}%
  \BibitemOpen
  \bibfield  {author} {\bibinfo {author} {\bibfnamefont {S.}~\bibnamefont
  {Schubotz}}, \bibinfo {author} {\bibfnamefont {Q.~A.}\ \bibnamefont
  {Besford}}, \bibinfo {author} {\bibfnamefont {S.}~\bibnamefont {Nazari}},
  \bibinfo {author} {\bibfnamefont {P.}~\bibnamefont {Uhlmann}}, \bibinfo
  {author} {\bibfnamefont {E.}~\bibnamefont {Bittrich}}, \bibinfo {author}
  {\bibfnamefont {J.-U.}\ \bibnamefont {Sommer}},\ and\ \bibinfo {author}
  {\bibfnamefont {G.~K.}\ \bibnamefont {Auernhammer}},\ }\bibfield  {title}
  {\bibinfo {title} {Influence of the atmosphere on the wettability of polymer
  brushes},\ }\href {https://doi.org/10.1021/acs.langmuir.2c03009} {\bibfield
  {journal} {\bibinfo  {journal} {Langmuir}\ }\textbf {\bibinfo {volume}
  {39}},\ \bibinfo {pages} {4872} (\bibinfo {year} {2023})}\BibitemShut
  {NoStop}%
\bibitem [{\citenamefont {Galuschko}\ and\ \citenamefont
  {Sommer}(2019)}]{Galuschko2019}%
  \BibitemOpen
  \bibfield  {author} {\bibinfo {author} {\bibfnamefont {A.}~\bibnamefont
  {Galuschko}}\ and\ \bibinfo {author} {\bibfnamefont {J.-U.}\ \bibnamefont
  {Sommer}},\ }\bibfield  {title} {\bibinfo {title} {Co-nonsolvency response of
  a polymer brush: A molecular dynamics study},\ }\href
  {https://doi.org/10.1021/acs.macromol.9b00569} {\bibfield  {journal}
  {\bibinfo  {journal} {Macromolecules}\ }\textbf {\bibinfo {volume} {52}},\
  \bibinfo {pages} {4120} (\bibinfo {year} {2019})}\BibitemShut {NoStop}%
\bibitem [{\citenamefont {Butt}\ \emph {et~al.}(2018)\citenamefont {Butt},
  \citenamefont {Berger}, \citenamefont {Steffen}, \citenamefont {Vollmer},\
  and\ \citenamefont {Weber}}]{Butt2018}%
  \BibitemOpen
  \bibfield  {author} {\bibinfo {author} {\bibfnamefont {H.-J.}\ \bibnamefont
  {Butt}}, \bibinfo {author} {\bibfnamefont {R.}~\bibnamefont {Berger}},
  \bibinfo {author} {\bibfnamefont {W.}~\bibnamefont {Steffen}}, \bibinfo
  {author} {\bibfnamefont {D.}~\bibnamefont {Vollmer}},\ and\ \bibinfo {author}
  {\bibfnamefont {S.~A.~L.}\ \bibnamefont {Weber}},\ }\bibfield  {title}
  {\bibinfo {title} {Adaptive wetting{\textemdash}adaptation in wetting},\
  }\href {https://doi.org/10.1021/acs.langmuir.8b01783} {\bibfield  {journal}
  {\bibinfo  {journal} {Langmuir}\ }\textbf {\bibinfo {volume} {34}},\ \bibinfo
  {pages} {11292} (\bibinfo {year} {2018})}\BibitemShut {NoStop}%
\bibitem [{\citenamefont {Ritsema~van Eck}\ \emph
  {et~al.}(2022{\natexlab{a}})\citenamefont {Ritsema~van Eck}, \citenamefont
  {Chiappisi},\ and\ \citenamefont {de~Beer}}]{Eck2022}%
  \BibitemOpen
  \bibfield  {author} {\bibinfo {author} {\bibfnamefont {G.~C.}\ \bibnamefont
  {Ritsema~van Eck}}, \bibinfo {author} {\bibfnamefont {L.}~\bibnamefont
  {Chiappisi}},\ and\ \bibinfo {author} {\bibfnamefont {S.}~\bibnamefont
  {de~Beer}},\ }\bibfield  {title} {\bibinfo {title} {Fundamentals and
  applications of polymer brushes in air},\ }\href
  {https://doi.org/10.1021/acsapm.1c01615} {\bibfield  {journal} {\bibinfo
  {journal} {ACS Applied Polymer Materials}\ }\textbf {\bibinfo {volume} {4}},\
  \bibinfo {pages} {3062} (\bibinfo {year} {2022}{\natexlab{a}})},\ \Eprint
  {https://arxiv.org/abs/https://doi.org/10.1021/acsapm.1c01615}
  {https://doi.org/10.1021/acsapm.1c01615} \BibitemShut {NoStop}%
\bibitem [{\citenamefont {Birshtein}\ and\ \citenamefont
  {Lyatskaya}(1994)}]{Birshtein1994}%
  \BibitemOpen
  \bibfield  {author} {\bibinfo {author} {\bibfnamefont {T.~M.}\ \bibnamefont
  {Birshtein}}\ and\ \bibinfo {author} {\bibfnamefont {Y.~V.}\ \bibnamefont
  {Lyatskaya}},\ }\bibfield  {title} {\bibinfo {title} {Theory of the
  collapse-stretching transition of a polymer brush in a mixed solvent},\
  }\href {https://doi.org/10.1021/ma00083a028} {\bibfield  {journal} {\bibinfo
  {journal} {Macromolecules}\ }\textbf {\bibinfo {volume} {27}},\ \bibinfo
  {pages} {1256} (\bibinfo {year} {1994})}\BibitemShut {NoStop}%
\bibitem [{\citenamefont {Ritsema~van Eck}\ \emph
  {et~al.}(2022{\natexlab{b}})\citenamefont {Ritsema~van Eck}, \citenamefont
  {Kiens}, \citenamefont {Veldscholte}, \citenamefont {Brió~Pérez},\ and\
  \citenamefont {de~Beer}}]{RitsemavanEck2022}%
  \BibitemOpen
  \bibfield  {author} {\bibinfo {author} {\bibfnamefont {G.~C.}\ \bibnamefont
  {Ritsema~van Eck}}, \bibinfo {author} {\bibfnamefont {E.~M.}\ \bibnamefont
  {Kiens}}, \bibinfo {author} {\bibfnamefont {L.~B.}\ \bibnamefont
  {Veldscholte}}, \bibinfo {author} {\bibfnamefont {M.}~\bibnamefont
  {Brió~Pérez}},\ and\ \bibinfo {author} {\bibfnamefont {S.}~\bibnamefont
  {de~Beer}},\ }\bibfield  {title} {\bibinfo {title} {Vapor swelling of polymer
  brushes compared to nongrafted films},\ }\href
  {https://doi.org/10.1021/acs.langmuir.2c01889} {\bibfield  {journal}
  {\bibinfo  {journal} {Langmuir}\ }\textbf {\bibinfo {volume} {38}},\ \bibinfo
  {pages} {13763} (\bibinfo {year} {2022}{\natexlab{b}})}\BibitemShut {NoStop}%
\bibitem [{\citenamefont {Besford}\ \emph {et~al.}(2022)\citenamefont
  {Besford}, \citenamefont {Merlitz}, \citenamefont {Schubotz}, \citenamefont
  {Yong}, \citenamefont {Chae}, \citenamefont {Schnepf}, \citenamefont {Weiss},
  \citenamefont {Auernhammer}, \citenamefont {Sommer}, \citenamefont
  {Uhlmann},\ and\ \citenamefont {Fery}}]{Besford2022}%
  \BibitemOpen
  \bibfield  {author} {\bibinfo {author} {\bibfnamefont {Q.~A.}\ \bibnamefont
  {Besford}}, \bibinfo {author} {\bibfnamefont {H.}~\bibnamefont {Merlitz}},
  \bibinfo {author} {\bibfnamefont {S.}~\bibnamefont {Schubotz}}, \bibinfo
  {author} {\bibfnamefont {H.}~\bibnamefont {Yong}}, \bibinfo {author}
  {\bibfnamefont {S.}~\bibnamefont {Chae}}, \bibinfo {author} {\bibfnamefont
  {M.~J.}\ \bibnamefont {Schnepf}}, \bibinfo {author} {\bibfnamefont
  {A.~C.~G.}\ \bibnamefont {Weiss}}, \bibinfo {author} {\bibfnamefont {G.~K.}\
  \bibnamefont {Auernhammer}}, \bibinfo {author} {\bibfnamefont {J.-U.}\
  \bibnamefont {Sommer}}, \bibinfo {author} {\bibfnamefont {P.}~\bibnamefont
  {Uhlmann}},\ and\ \bibinfo {author} {\bibfnamefont {A.}~\bibnamefont
  {Fery}},\ }\bibfield  {title} {\bibinfo {title} {Mechanofluorescent polymer
  brush surfaces that spatially resolve surface solvation},\ }\href
  {https://doi.org/10.1021/acsnano.2c00277} {\bibfield  {journal} {\bibinfo
  {journal} {ACS Nano}\ }\textbf {\bibinfo {volume} {16}},\ \bibinfo {pages}
  {3383} (\bibinfo {year} {2022})},\ \bibinfo {note} {pMID: 35112848},\ \Eprint
  {https://arxiv.org/abs/https://doi.org/10.1021/acsnano.2c00277}
  {https://doi.org/10.1021/acsnano.2c00277} \BibitemShut {NoStop}%
\end{thebibliography}%

\end{document}


\title{Supplemental Material for: Positive Feedback Drives Sharp Swelling of Polymer Brushes near Saturation}

\author{Simon Schubotz}
\email{schubotz@ipfdd.de}
\affiliation
{Leibniz-Institut für Polymerforschung Dresden e.V, Hohe Stra\ss e 6, Dresden 01069, Germany}
\affiliation
{Technische Universität Dresden, Helmholtztra\ss e 10, Dresden 01062, Germany}

\author{Eva Bittrich}
\affiliation
{Leibniz-Institut für Polymerforschung Dresden e.V, Hohe Stra\ss e 6, Dresden 01069, Germany}

\author{Holger Merlitz}
\affiliation
{Leibniz-Institut für Polymerforschung Dresden e.V, Hohe Stra\ss e 6, Dresden 01069, Germany}

\author{Quinn A. Besford}
\affiliation
{Leibniz-Institut für Polymerforschung Dresden e.V, Hohe Stra\ss e 6, Dresden 01069, Germany}

\author{Petra Uhlmann}
\affiliation
{Leibniz-Institut für Polymerforschung Dresden e.V, Hohe Stra\ss e 6, Dresden 01069, Germany}

\author{Jens-Uwe Sommer}
\affiliation
{Leibniz-Institut für Polymerforschung Dresden e.V, Hohe Stra\ss e 6, Dresden 01069, Germany}
\affiliation
{Institute for Theoretical Physics, Technische Universität Dresden, 01069 Dresden, Germany}

\author{Günter K. Auernhammer}
\email{auernhammer@ipfdd.de}
\affiliation
{Leibniz-Institut für Polymerforschung Dresden e.V, Hohe Stra\ss e 6, Dresden 01069, Germany}

\date{\today}
\maketitle

\renewcommand{\theequation}{S\arabic{equation}}
\renewcommand{\thefigure}{S\arabic{figure}}
\renewcommand{\thetable}{S\arabic{table}}
\renewcommand{\thesection}{S\arabic{section}}
\renewcommand{\thepage}{S\arabic{page}}

\setcounter{equation}{0}
\setcounter{figure}{0}
\setcounter{table}{0}
\setcounter{section}{0}
\setcounter{page}{1}


\section{\label{app:Details_MM}Futher details on Materials and Methods}

\subsection{Data Management}
Data organization and analysis were performed using a custom electronic laboratory notebook (ELN) designed for rapid, partially automated processing of large datasets~\cite{schubotz2025, SchubotzCode2022}. The ELN's detailed documentation of the experimental history for each sample was crucial for ensuring the reproducibility of our measurements, particularly when investigating samples that experience a memory effect.

\subsection{Polymer Brush Preparation}

\subsubsection{Materials}
Monocarboxy-terminated poly(\textit{N}-isopropylacrylamide) (PNIPAAm-COOH, $M_n = 61,000$~g/mol, $M_w/M_n = 1.4$) was synthesized according to a previously published procedure~\cite{Rauch2012}. Poly(glycidyl methacrylate) (PGMA, $M_n = 15,000$~g/mol, $M_w/M_n = 1.6$) was purchased from Polymer Source, Inc. The substrates used were p-type silicon wafers (100) featuring a native SiO$_2$ layer of approximately \SI{2}{\nano\meter} (Si-Mat, Landberg, Germany). For substrate cleaning and polymer synthesis, the following solvents were used as received: absolute ethanol (VMR Chemicals, Germany) for wafer treatment, and chloroform (CHCl$_3$) and tetrahydrofuran (THF) (Fisher Chemicals, U.S.; ACROS Organics, Belgium) for solution preparation and polymer extraction.

\subsubsection{Grafting-To Procedure}
The PNIPAAm brushes were prepared using a two-step "grafting-to" procedure~\cite{Zdyrko2006, Iyer2003}.

Silicon wafers were first thoroughly cleaned and then activated via plasma oxidation for \SI{1}{\minute} at \SI{100}{\watt} in a 440-G Plasma System (Technics Plasma GmbH, Germany). Immediately following activation, a \SI{0.02}{wt\percent} solution of PGMA in chloroform was spin-coated onto the wafers (2000 rpm, 1000 rpm/s acceleration, \SI{10}{s}). The substrates were then annealed in a vacuum oven at \SI{100}{\celsius} for \SI{20}{\minute}. This process formed a PGMA anchoring layer with a thickness of \SI{2.5 \pm 0.5}{\nano\meter} via the reaction of PGMA's epoxy groups with the surface silanol groups.

In the second step, a \SI{1}{wt\percent} solution of PNIPAAm-COOH in THF was spin-coated onto the PGMA-functionalized surface (3000 rpm, 1000 rpm/s acceleration, \SI{15}{s}). The samples were subsequently annealed at \SI{173}{\celsius} for \SI{16}{\hour} under vacuum, during which the carboxyl groups of the PNIPAAm chains formed ester bonds with the remaining epoxy groups of the anchoring layer. To remove any non-covalently bound polymer, the samples were incubated in deionized water for \SI{8}{\hour}. A final rinse with absolute ethanol followed by drying under a nitrogen stream yielded the finished polymer brushes.

\subsubsection{Brush Characterization}
The resulting PNIPAAm brushes had a dry thickness of $14 \pm 1$~nm, which corresponds to a grafting density of $\sigma \approx 0.15$~chains/nm$^2$. To confirm that the chains are in a true brush regime, we verified that the ratio of the grafting distance, $s$, to the radius of gyration, $R_G$, satisfies the condition $s / (2 R_G) \ll 1$~\cite{Bittrich2012, Xue2011}. For our system, this ratio was calculated to be approximately 0.1, confirming a sufficiently high grafting density. All experiments reported here were conducted on two identically prepared samples.

\subsection{Ellipsometry Setup and Methodology}

\subsubsection{Instrumentation and Optical Model}
Spectroscopic ellipsometry measurements were performed with an alpha-SE ellipsometer (J.A. Woollam Co., Inc., Lincoln, NE, USA), operating over a wavelength range of \SIrange{400}{900}{\nano\meter} at a \SI{70}{\degree} angle of incidence. Data were collected and analyzed using the CompleteEASE software (J.A. Woollam Co. Inc.).

The brush thickness was determined by fitting the ellipsometric angles to a four-layer optical box model comprising the silicon substrate, a native SiO$_2$ layer, the PGMA anchoring layer, and the PNIPAAm brush layer. Optical constants for Si and SiO$_2$ were taken from the software's database. Following the literature, the PGMA layer was modeled with a fixed thickness of \SI{2}{\nano\meter} and a Cauchy dispersion ($n(\lambda) = 1.522 + 0.005/\lambda^2$)~\cite{Bittrich2012}. The optical dispersion of the PNIPAAm layer was described by a Cauchy model ($n(\lambda) = 1.421 + 0.006/\lambda^2$, where $\lambda$ is the wavelength in \unit{\micro\meter}), with only its thickness as a free fitting parameter~\cite{Bittrich2012,Rauch2012}. We have previously confirmed that atmospheric changes alone introduce apparent thickness variations of less than \SI{0.5}{\nano\meter}~\cite{Schubotz2023}.

\subsubsection{Custom Sample Cell and Measurements in Liquid Water}
\begin{figure*}
    \centering
    \begin{subfigure}[b]{0.45\textwidth}
        \centering
        \includegraphics[width=0.7\linewidth]{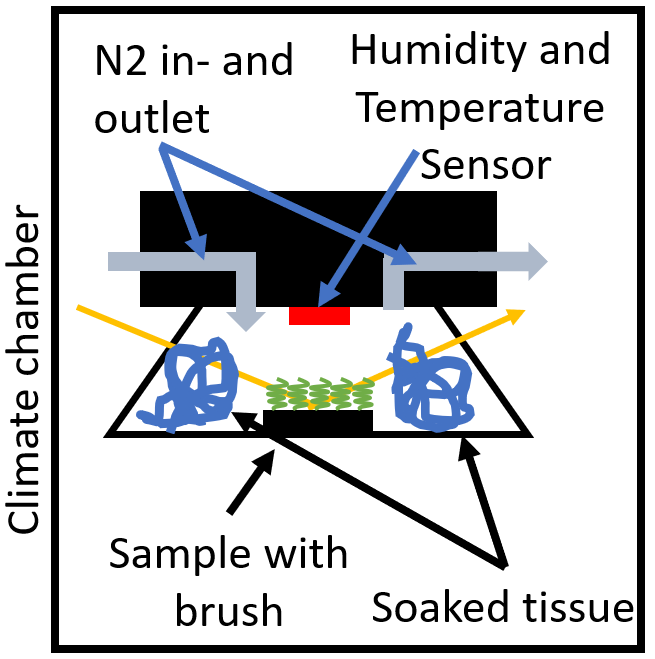}
        \caption{Schematic of the custom cell}
        \label{fig:setup_scheme_SI}
    \end{subfigure}
    \hfill
    \begin{subfigure}[b]{0.45\textwidth}
        \centering
        \includegraphics[width=0.9\linewidth]{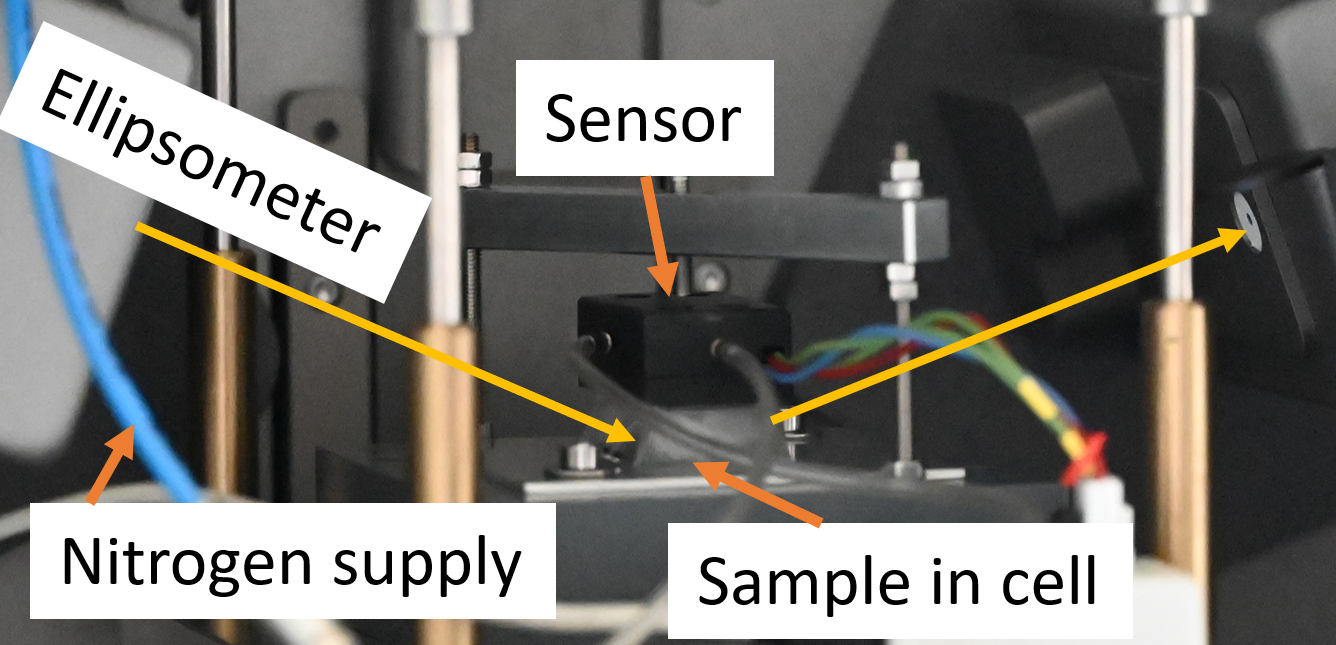}
        \caption{Photograph of the setup}
        \label{fig:setup_photo_SI}
    \end{subfigure}
	\caption{
	    (a) Schematic of the custom-built sample cell used for ellipsometry measurements. (b) Photograph of the experimental setup inside the climate chamber, shown in its configuration for vapor-phase measurements. The sample cell rests on a plastic block with a metal plate on top. The 3D-printed lid seals the cell and integrates the sensor and gas ports, enabling precise environmental control during the experiments.
	}
	\label{fig:setup_combined_SI}
\end{figure*}

All measurements were conducted within a custom-built cell ($V < \SI{10}{\milli\liter}$) with \SI{70}{\degree} angled walls to prevent beam distortion (\autoref{fig:setup_combined_SI}). To measure the brush thickness in liquid, this cell was filled with deionized water and mounted on a Peltier device for active temperature control. The water temperature was varied in discrete steps across the range of \SIrange{10}{35}{\degreeCelsius}. At each setpoint, the system was allowed to fully equilibrate—confirmed by monitoring both the water temperature and the brush thickness until they stabilized—before an ellipsometry measurement was taken to determine the equilibrium swollen thickness.

For all vapor-phase experiments, the entire ellipsometer setup, including the sample cell, was placed inside a climate chamber to ensure a stable thermal environment. The cell rested on a plastic block with a metal plate on top, as shown in \autoref{fig:setup_photo_SI}. The cell was sealed with a 3D-printed lid that integrated a humidity/temperature sensor (HYT 939, IST AG), a gas inlet, and a gas outlet. The atmosphere within the cell was controlled using two distinct methods:

\paragraph{Intermediate Humidity (\SIrange{10}{90}{\percent}~RH):}
A gas-mixing system was used, following the procedure described in Ref.~\cite{Schubotz2023}. A stream of nitrogen was saturated with water by bubbling it through a gas wash bottle (DURAN Group GmbH, Wertheim, Germany). This saturated stream was then mixed with a parallel stream of pure, dry nitrogen. The flow rates of both streams were regulated by mass flow controllers (ANALYT-MTC Me\ss technik GmbH, Müllheim, Germany) with a maximum flow of \SI{0.5}{\liter\per\minute} per channel, allowing for precise adjustment of the final humidity sent to the cell.

\paragraph{High Humidity Regime (up to \SI{100}{\percent}~RH):}
A saturated atmosphere was first created inside the cell using water-soaked tissue (\autoref{fig:setup_scheme_SI}). The relative humidity was then precisely tuned down from \SI{100}{\percent} by introducing a controlled, low flow rate of dry nitrogen using the same mass flow controller system.

In addition to the in-cell HYT 939 sensor, the ambient temperature and humidity inside the climate chamber were continuously monitored with a P790 instrument (Dostmann Electronic GmbH, Germany).

\subsection{Molecular Dynamics Simulations}

\subsubsection{Model and Potentials}
Molecular dynamics (MD) simulations were performed using the LAMMPS software package. We employed a coarse-grained, bead-spring model for the PNiPAAm brush, where each bead represents a Kuhn segment, and chains are formed by connecting beads with finite extensible nonlinear elastic (FENE) bonds~\cite{Galuschko2019}.

All non-bonded interactions between beads (polymer, water, cosolvent) were described by a Lennard-Jones (LJ) potential with a cutoff radius of $2.5d$:
\begin{equation}
    V_\mathrm{LJ}(r) = 4\epsilon \left[ \left(\frac{d}{r}\right)^{12} - \left(\frac{d}{r}\right)^6 \right]
\end{equation}
Here, $d$ is the bead diameter and $\epsilon$ is the interaction strength, which was set depending on the bead types. The key parameters used to model different solvent qualities were the polymer-water interaction, which was set to $\epsilon_\mathrm{pw} = 1.4$ or $\epsilon_\mathrm{pw} = 1.6$, while other parameters were kept constant: $\epsilon_\mathrm{pp} = 1.2$ (polymer-polymer) and $\epsilon_\mathrm{ww} = 1.1$ (water-water).

The simulation box dimensions were approximately $38d \times 38d \times 80d$, with periodic boundary conditions in the x- and y-directions. The brush was grafted to a flat, impenetrable substrate at the bottom ($z=0$), which was modeled as an integrated 9--3 LJ wall potential.

\subsubsection{Simulation Protocol}
The system was equilibrated using a multi-stage protocol. First, 36 monodisperse polymer chains (N=120 beads) were grown from the substrate via a random walk. An initial swelling stage with purely repulsive potentials (cutoff at $1.12d$) was used to remove bead overlaps and achieve a well-swollen initial state. The grafted beads were anchored by setting their forces to zero.

Second, the system was equilibrated in the canonical (NVT) ensemble, maintaining a constant Number of particles, Volume, and Temperature, for $1 \times 10^6$ timesteps using a Langevin thermostat until the brush's center of mass stabilized.

Third, to control the vapor phase, the system was transitioned to a grand canonical ($\mu$VT) ensemble. The chemical potential of the solvent was set using a Monte Carlo algorithm for $5 \times 10^6$ steps to introduce solvent beads into the vapor phase while the brush was temporarily immobilized.

Finally, the brush was released, and the full attractive LJ potentials (cutoff at $2.5d$) were activated. The entire system was then equilibrated for an additional $1 \times 10^7$ timesteps in the NVT ensemble until a final, stable swollen state was reached. The relative humidity for a given chemical potential was determined by relating the pressure in the solvated brush system to the pressure at the liquid-vapor coexistence point in a separate calibration simulation. The final equilibrium brush height was calculated as twice the vertical position of the center of mass of all polymer beads.

\subsection{\label{app:correction_of_humidity_data}Processing and Correction of Swelling Isotherm Data}

\begin{figure*}
    \centering
    \begin{subfigure}[c]{0.4\textwidth}
        \includegraphics[width=\linewidth]{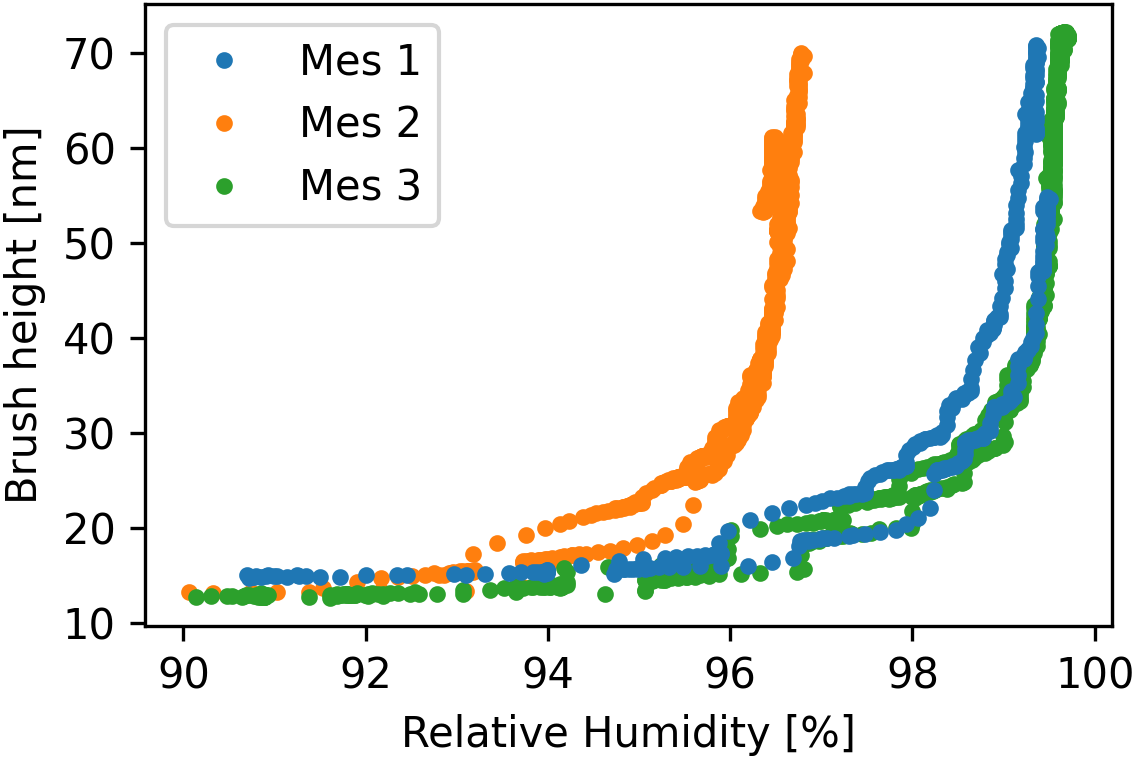}
        \caption{Before corrections}
        \label{fig:thickness_brush_12_no_corr}
    \end{subfigure}
    \hfill
    \begin{subfigure}[c]{0.4\textwidth}
        \includegraphics[width=\linewidth]{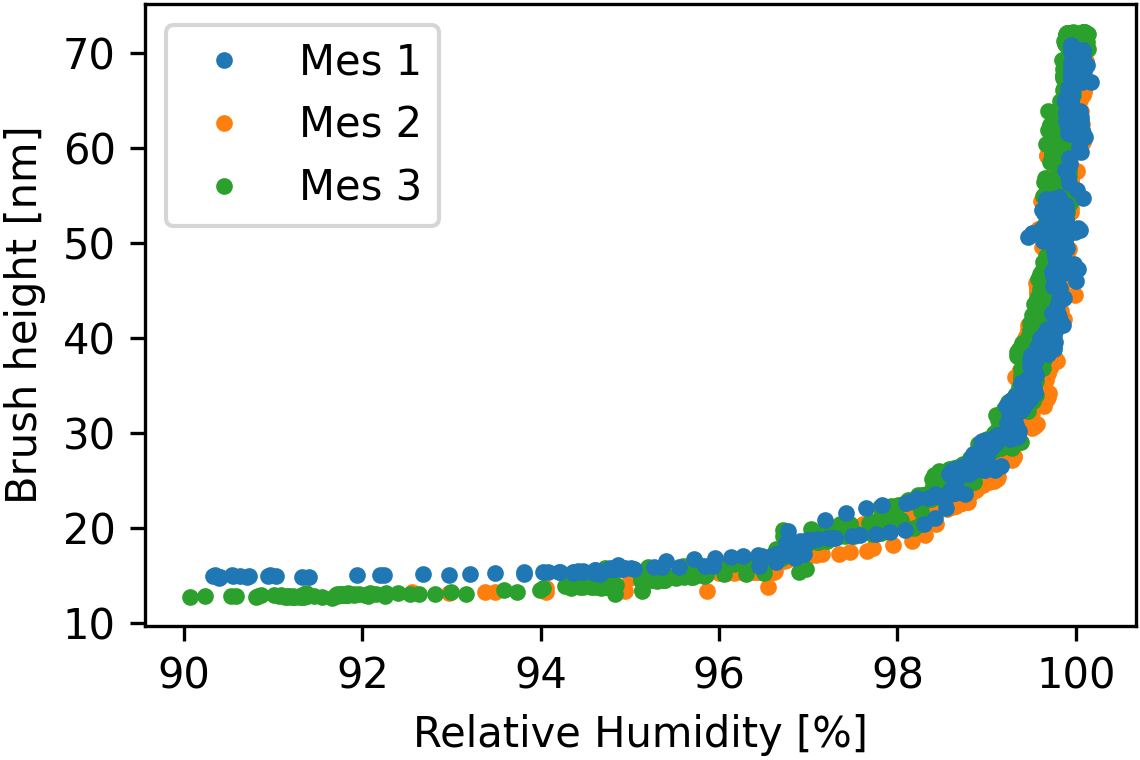}
        \caption{After corrections}
        \label{fig:thickness_only}
    \end{subfigure}
    \caption{
        Comparison of brush swelling isotherms plotted against relative humidity (a) before and (b) after the full correction procedure. The raw data in (a) shows significant hysteresis and incorrect saturation points. The final corrected data in (b) is free from these artifacts, showing a collapse of the hysteresis loops and a consistent saturation point at 100\%~RH.
    }
    \label{fig:elli_mes}
\end{figure*}

The raw experimental data of brush thickness versus measured relative humidity (RH) required a two-step correction procedure to account for instrumental artifacts. Figure~\ref{fig:elli_mes} provides a direct comparison of the swelling isotherms before (a) and after (b) these corrections were applied.

\subsubsection{Correction for Sensor Drift and Calibration}

The raw data (\autoref{fig:thickness_brush_12_no_corr}) consistently showed a maximum swelling at a measured RH value below \SI{100}{\percent}, an artifact attributed to a calibration error in the humidity sensor. We corrected this by using the onset of condensation as a reference for 100\%~RH. The condensation point was identified as the moment the optical model produced unphysical behavior (a simultaneous increase in both thickness and fitted refractive index). The entire RH axis for each measurement was then shifted to align this point with exactly \SI{100}{\percent}~RH. The validity of this linear shift is supported by the symmetric response of the sensor during the humidification and dehumidification cycles. We also observed that prolonged exposure to high saturation caused the sensor to drift and prematurely report \SI{100}{\percent}~RH on subsequent runs. Sensors that had drifted too much were replaced.

\subsubsection{Correction for Thermal Hysteresis}

\begin{figure}
    \centering
    \includegraphics[width=0.7\linewidth]{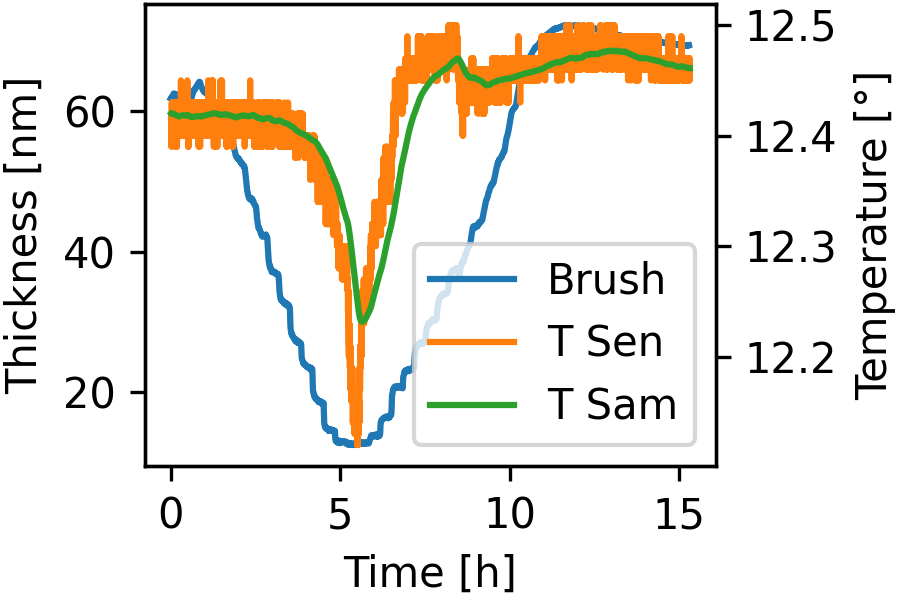}
    \caption{
        Example of the thermal lag correction. The raw temperature data measured by the sensor (orange line) is used to calculate the estimated actual sample temperature (green line) using the first-order lag model in Eq.~\eqref{eq:thermal_lag}.
    }
    \label{fig:example_high_tmp}
\end{figure}

After the initial RH-axis correction, a hysteresis loop was still visible between the swelling and deswelling curves. We attribute this to a thermal mismatch between the sample and the sensor, which are not perfectly isothermal. To correct for this, we modeled the true sample temperature, $T_s$, as lagging behind the sensor temperature, $T_\text{sensor}$, due to thermal inertia. The sample temperature was estimated iteratively by modeling it as a first-order lag process:
\begin{equation}
    T_s(t_n) = T_s(t_{n-1}) + k\left(T_{\text{sensor}}(t_n) - T_s(t_{n-1})\right) \Delta t
    \label{eq:thermal_lag}
\end{equation}
where $t_n$ is the time of the n-th measurement, $\Delta t = t_n - t_{n-1}$ is the time step between consecutive measurements. A dataset-dependent rate constant, $k$, was chosen in the range of \SIrange{0.0006}{0.002}{\per\second} to best remove the thermal hysteresis. This optimization was necessary for high precision; for instance, using a single constant value of $k=\SI{0.002}{\per\second}$ for all datasets would leave a residual gap between the swelling and deswelling curves of approximately \SI{0.2}{\percent} relative humidity in the highly swollen state. This model allows the calculated sample temperature to gradually ``catch up'' to the measured sensor temperature, as illustrated in \autoref{fig:example_high_tmp}.

Using this calculated sample temperature, the relative humidity at the sample was recalculated for each time point. The final, fully corrected swelling isotherms, which are free from both sensor drift and thermal hysteresis, are presented in \autoref{fig:thickness_only}.

\subsubsection{Alignment of Datasets from Different Humidity Regimes}
The full swelling isotherm presented in the main text is a composite of data from two different experimental methods: one for the intermediate humidity range (\SIrange{10}{90}{\percent}~RH) using a gas-mixing system, and another for the high humidity regime (\SIrange{80}{100}{\percent}~RH) using an in-cell vapor source. Because the sample was sometimes removed and remounted between these two types of experiments, minor variations in the exact measurement spot on the sample could occur.

This occasionally resulted in a small vertical offset (up to \SI{2.4}{\nano\meter}) in the measured brush thickness between the two datasets in their overlapping humidity range. To create a single, continuous isotherm, a small vertical shift was applied to one of the datasets to ensure they matched seamlessly in the 80--90\%~RH region. This alignment procedure corrects for the slight variation in measurement location and does not affect the shape or the conclusions drawn from the swelling curves.

\section{\label{app:determination_cond_onset}Determination of the 100\% RH Point}

\begin{figure*}
	\centering
	\begin{subfigure}[c]{0.4\textwidth}	
    	\includegraphics[width=\linewidth]{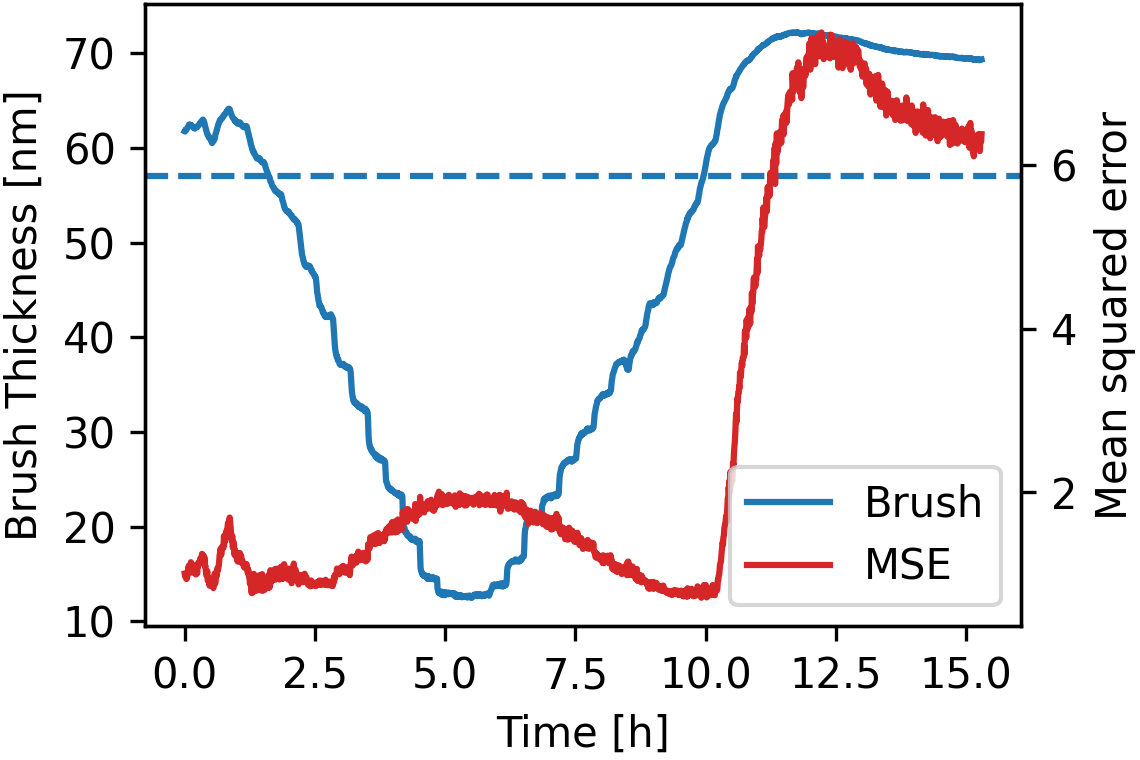}
    	\caption{Mean Squared Error (MSE) of the fit}
    	\label{fig:example_high_mse}
	\end{subfigure}
    \hfill
    \begin{subfigure}[c]{0.4\textwidth}	
    	\includegraphics[width=\linewidth]{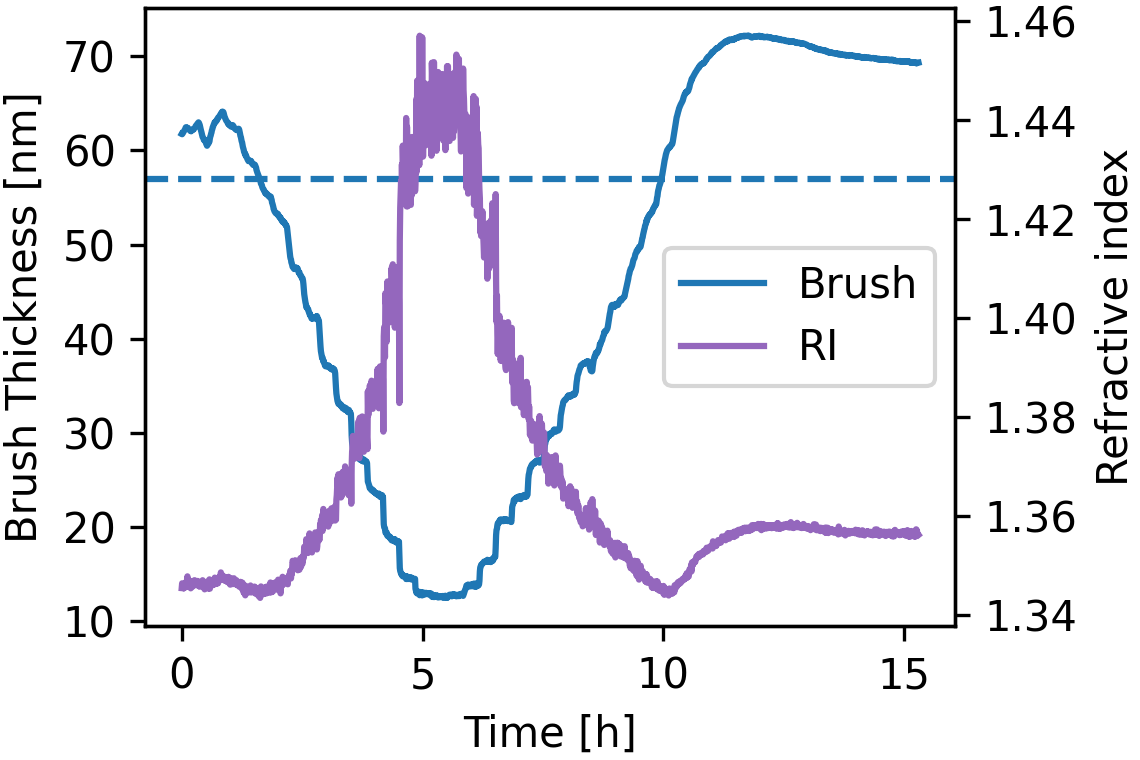}
    	\caption{Fitted Refractive Index of the brush layer}
    	\label{fig:example_high_n}
	\end{subfigure}
	\caption{
	    Determination of the 100\% RH point from the optical model fit.
	    (a) The Mean Squared Error (MSE) remains low during swelling but shows a sharp and sustained increase shortly after $t \approx \SI{10}{\hour}$, confirming the presence of condensed droplets via light scattering.
	    (b) The fitted refractive index of the brush layer decreases as the brush swells. The onset of condensation is identified as the earlier, more sensitive indicator: the unphysical inversion point (here, at $t \approx \SI{10}{\hour}$) where the refractive index begins to increase despite the continued increase in apparent thickness. This inversion point is defined as the 100\%~RH cutoff for our analysis.
	}
	\label{fig:condensation_onset}
\end{figure*}

To accurately analyze the swelling isotherm, it was crucial to precisely identify the point of 100\% relative humidity, beyond which ellipsometry data becomes false. We identified this point by monitoring two key outputs from the optical model fit: the fitted refractive index of the brush layer and the mean squared error (MSE), as shown in \autoref{fig:condensation_onset}.

Up to approximately \SI{10}{\hour}, both indicators exhibit physically reasonable behavior. The refractive index (\autoref{fig:example_high_n}) correctly decreases as the brush swells with increasing water content, and the MSE (\autoref{fig:example_high_mse}) remains low, indicating a good model fit.

The first sign of condensation is an unphysical inversion in the refractive index trend. At approximately the 10-hour mark, while the apparent thickness continues to increase, the fitted refractive index also begins to increase. This is physically inconsistent, as further water uptake must lead to a decrease in the layer's average refractive index. We therefore define this inversion point as the precise onset of 100\%~RH and the cutoff for our thermodynamic data. Shortly after this point, the MSE also shows a sharp and sustained increase, which corresponds to visible light scattering from condensed microdroplets and confirms the failure of the simple optical model. All data recorded after the initial refractive index inversion are excluded from analysis.

A closer inspection of the refractive index data (\autoref{fig:example_high_n}) occasionally reveals minor fluctuations or slight deviations from the expected trend within the first two hours of the experiment. These subtle features, which are not present in every run and depend on the precise starting thickness, could be interpreted as signs of minor, transient condensation. However, these indications are not as clear or sustained as the unambiguous, unphysical inversion observed after $t \approx \SI{10}{\hour}$. Therefore, while we acknowledge that this might introduce a small uncertainty in the initial part of the isotherm, we do not focus on these transient effects in our primary analysis. A more sophisticated experimental protocol, perhaps involving a feedback loop between the measured brush thickness and the nitrogen flow rate, would be required to characterize this highly sensitive regime near saturation with even greater detail.

\section{Application of the Adaptive Wetting Model}

The dynamic wetting data presented in the main text (Figure 4a) were quantitatively analyzed using the adaptive wetting model from Butt~et~al.~\cite{Butt2018}. This model provides an expression for the dynamic advancing contact angle, $\theta_a$, as a function of contact line velocity, $U$:
\begin{equation}
\cos(\theta_a(U)) = \cos(\theta_\infty) - \frac{\Delta\gamma_\mathrm{SL}}{\gamma_\mathrm{LG}} \exp\left(-\frac{\nu_\mathrm{SL}}{U}\right)
\label{eq:adaptive_wetting_si}
\end{equation}
Here, $\theta_\infty$ is the fully relaxed contact angle at $U \to 0$, $\gamma_\mathrm{LG}$ is the liquid-vapor surface tension, and $\nu_\mathrm{SL} = l_\mathrm{SL} / \tau_\mathrm{SL}$ is the characteristic velocity of surface adaptation. The key fitting parameter, $\Delta\gamma_\mathrm{SL} = \gamma_\mathrm{SL}^0 - \gamma_\mathrm{SL}^\infty$, represents the magnitude of the total change in solid-liquid interfacial energy as the surface relaxes from its initial state ($\gamma_\mathrm{SL}^0$) to its final equilibrium state ($\gamma_\mathrm{SL}^\infty$).

\begin{figure}
	\centering
	\includegraphics[width=0.8\linewidth]{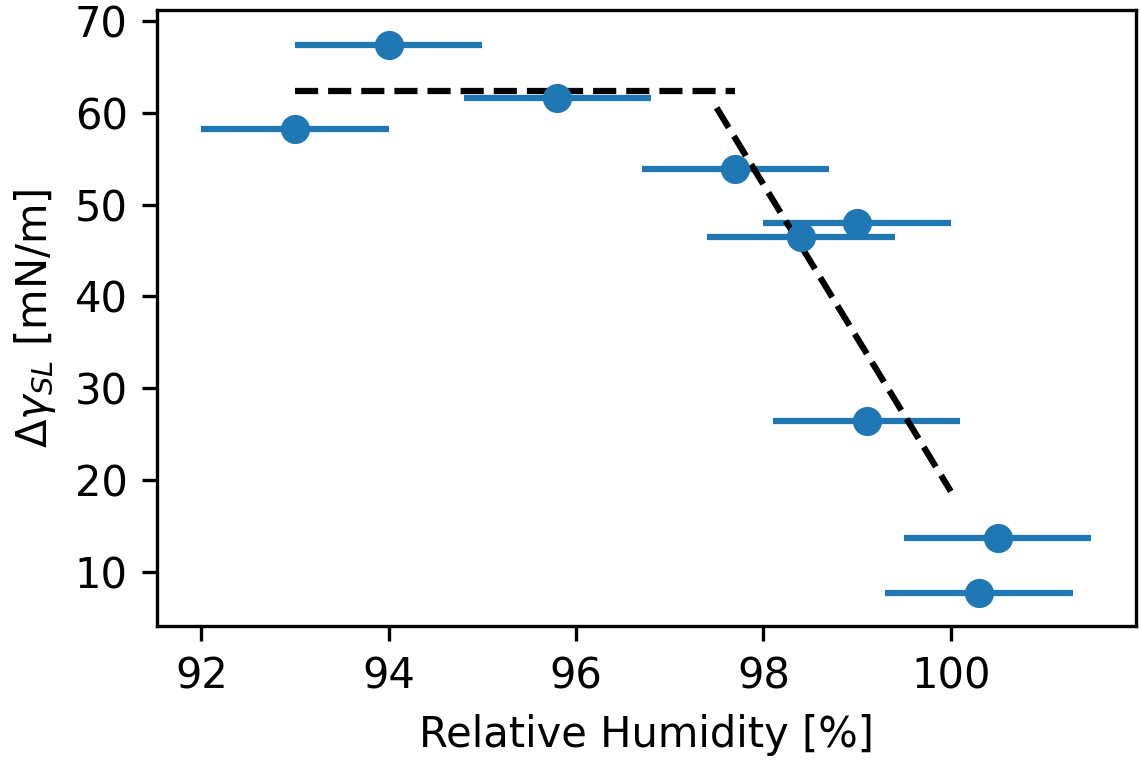}
	\caption{
	    The magnitude of the interfacial energy change, $\Delta\gamma_\mathrm{SL}$, as a function of ambient relative humidity. Values are extracted from the global fit of the adaptive wetting model (Eq.~\eqref{eq:adaptive_wetting_si}) to the dynamic contact angle data shown in the main text. The plot shows that $\Delta\gamma_\mathrm{SL}$, which quantifies the energetic scale of the surface adaptation, is large in a dry environment but drops sharply above 98\%~RH. This behavior directly mirrors the sharp increase in the static swelling isotherm, providing the quantitative link between the equilibrium pre-swelling of the brush and its dynamic wetting response. The dashed line is a guide to the eye.
	}
	\label{fig:delta_gamma_sl}
\end{figure}

The solid lines shown in Figure 4a of the main text are the result of a simultaneous global fit of Eq.~\eqref{eq:adaptive_wetting_si} to all humidity-dependent data curves. For this global fit, the fully relaxed contact angle was fixed to the experimentally observed low-velocity limit of $\theta_\infty = 70^\circ$. A single, global value for the characteristic velocity, $\nu_\mathrm{SL} \approx \SI{0.14}{mm/s}$, was fitted across all datasets. The key parameter, $\Delta\gamma_\mathrm{SL}$, was allowed to vary as a free parameter for each ambient humidity condition. The resulting values for $\Delta\gamma_\mathrm{SL}$ versus ambient relative humidity are the data points plotted in \autoref{fig:delta_gamma_sl}.

\section{Details on the Meanfield Model}

\subsection{\label{app:derev_f}Derivation of the Grand Potential Free Energy per Monomer}

Here, we provide a derivation of the grand potential free energy per monomer, $f(\varphi)$, as presented in the main text. The derivation begins with the standard Helmholtz free energy for a polymer brush, $F_\text{brush}$, following the approach used by Birshtein et al.~\cite{Eck2022, Birshtein1994}.

The Helmholtz free energy of the brush, normalized by $k_\mathrm{B}T$, includes terms for solvent entropy, polymer-solvent interactions, and the elastic energy of the stretched chains (cf. Eq. 4 in Ref.~\cite{RitsemavanEck2022}):
\begin{equation}
    \frac{F_\text{brush}}{k_\mathrm{B}T} = n_s \ln(\varphi) + \chi n_s (1-\varphi) + n_p \frac{3N \rho_g^2}{2(1-\varphi)^2}
    \label{eq:SI_F_brush_final}
\end{equation}
where $n_s$ is the number of solvent molecules, $n_p$ is the number of polymer chains, $N$ is the number of monomers per chain, $\varphi$ is the solvent volume fraction, $\chi$ is the Flory-Huggins parameter, and $\rho_g$ is the dimensionless grafting density (details on the grafting density can be found in Appendix~\ref{app:grafting_den}).

To analyze the system in equilibrium with a vapor phase at a constant chemical potential, $\mu$, we consider the grand potential, $G = F_\text{brush} - \mu n_s$. We then define the free energy per monomer, $f$, by normalizing $G$ by the total number of monomers, $N_\text{mono} = n_p N$:
\begin{equation}
    f = \frac{G}{N_\text{mono}} = \frac{F_\text{brush}}{N_\text{mono}} - \frac{\mu n_s}{N_\text{mono}}
\end{equation}
To perform this normalization, we use the identity relating the number of solvent molecules to the total number of monomers and the solvent volume fraction: $n_s = N_\text{mono} \frac{\varphi}{1-\varphi}$. Substituting this and $n_p N = N_\text{mono}$ into the expression for $f$ along with Eq.~\eqref{eq:SI_F_brush_final} yields:
%
%
%
\begin{equation}
    f(\varphi) = \frac{3}{2} \frac{\rho_g^2}{(1-\varphi)^2} + \frac{\varphi}{1-\varphi} \ln(\varphi) + \chi(\varphi) \cdot \varphi - \frac{\mu \cdot \varphi}{1-\varphi}
    \label{for:free_f_si}
\end{equation}
This corresponds to Eq.~(1) in the main text.

\subsection{\label{app:solv_DGL}Solving differential equation}

The free energy $f$ is given in \autoref{for:free_f_si}.
For convenience, we separate this into a term $f_0(\varphi, \mu)$ that is independent of the interaction parameter $\chi$, and an interaction term $\eta(\varphi) = \chi(\varphi)\varphi$.

The equilibrium state is found by minimizing the free energy, which requires setting its derivative with respect to $\varphi$ equal to zero. Differentiating $f$ yields the equilibrium condition:
\begin{align}
    \frac{\partial f}{\partial \varphi} &= \left[ \frac{3\,\rho_g^2}{(1-\varphi)^3} + \frac{\ln(\varphi)}{(1-\varphi)^2} + \frac{1}{1-\varphi} - \frac{\mu}{(1-\varphi)^2} \right] \nonumber \\
    &\qquad + \chi(\varphi) + \varphi\,\chi'(\varphi) = 0.
    \label{eq:equilibrium_full}
\end{align}
We can simplify this by defining the bracketed terms, which represent $\partial f_0 / \partial \varphi$, as $g(\varphi, \mu)$. This allows us to express the equilibrium condition as a first-order linear differential equation for the concentration-dependent interaction parameter $\chi(\varphi)$:
\begin{align}
	\frac{\partial f_0}{\partial \varphi} = g(\varphi,\mu) = -\chi(\varphi) - \varphi\,\chi'(\varphi). \label{eq:diff_eq_si}
\end{align}
To solve this equation, we use the method of variation of parameters. First, we solve the homogeneous part of the equation, $\chi(\varphi) + \varphi\chi'(\varphi) = 0$. The general solution is $\chi_h(\varphi) = C/\varphi$, where $C$ is an integration constant.

Next, we seek a particular solution of the form $\chi(\varphi) = a(\varphi)\chi_h(\varphi)$. Setting $C=1$ for convenience gives $\chi(\varphi) = a(\varphi)/\varphi$. Substituting this form into the full differential equation \eqref{eq:diff_eq_si} results in:
\begin{align}
    g(\varphi,\mu) &= -\frac{a(\varphi)}{\varphi} - \varphi \cdot \frac{d}{d\varphi}\left(\frac{a(\varphi)}{\varphi}\right) \nonumber \\
    &= -\frac{a(\varphi)}{\varphi} - \varphi \left( \frac{a'(\varphi)}{\varphi} - \frac{a(\varphi)}{\varphi^2} \right) \nonumber \\
    &= -\frac{a(\varphi)}{\varphi} - a'(\varphi) + \frac{a(\varphi)}{\varphi} \nonumber \\
    &= -a'(\varphi).
\end{align}
Since we defined $g(\varphi,\mu) = \partial f_0 / \partial \varphi$, we have $\partial f_0 / \partial \varphi = -a'(\varphi)$. Integrating with respect to $\varphi$ yields:
\begin{align}
    \int \frac{\partial f_0}{\partial \varphi} d\varphi = \int -a'(\varphi) d\varphi \, \Rightarrow \, f_0(\varphi) = -a(\varphi) + c,
\end{align}
where $c$ is the integration constant. We can determine the constant by setting a physical boundary condition. A convenient choice is to define the interaction parameter relative to the state at $\varphi=0$, which sets $c = f_0(0)$. This gives $a(\varphi) = -f_0(\varphi) + f_0(0)$.

Substituting this back into our expression for $\chi(\varphi)$, we obtain the final solution:
\begin{align}
    \chi(\varphi)=\frac{1}{\varphi}\Bigl(f_0(0)-f_0(\varphi)\Bigr). \label{eq:chi_solution_si}
\end{align}
As a consistency check, we can verify that this solution satisfies the original differential equation. Differentiating our expression for $\chi(\varphi)$ with respect to $\varphi$ gives:
\begin{align}
\chi'(\varphi) &= -\frac{1}{\varphi^2}\Bigl(f_0(0)-f_0(\varphi)\Bigr) - \frac{1}{\varphi}\frac{\partial f_0}{\partial\varphi} \nonumber \\ 
&= -\frac{\chi(\varphi)}{\varphi} - \frac{1}{\varphi}\frac{\partial f_0}{\partial\varphi}.
\end{align}
Rearranging for $\partial f_0 / \partial \varphi$, we find:
\begin{align}
    \frac{\partial f_0}{\partial \varphi} = -\chi(\varphi) - \varphi\,\chi'(\varphi),
\end{align}
which is the differential equation we set out to solve, confirming the correctness of our solution.

\subsection{\label{app:closer_look}Detailed Analysis of the Temperature Dependence of $\chi(\varphi)$}

As stated in the main text, while the calculated $\chi_\mathrm{class}(\varphi)$ curves appear visually similar across different temperatures, the underlying $\chi(\varphi)$ functions are distinct. This section provides a detailed analysis of these differences and explains their significance.

\begin{figure*}[!htbp]
	\centering
	\begin{subfigure}[b]{0.45\textwidth}
		\includegraphics[width=\linewidth]{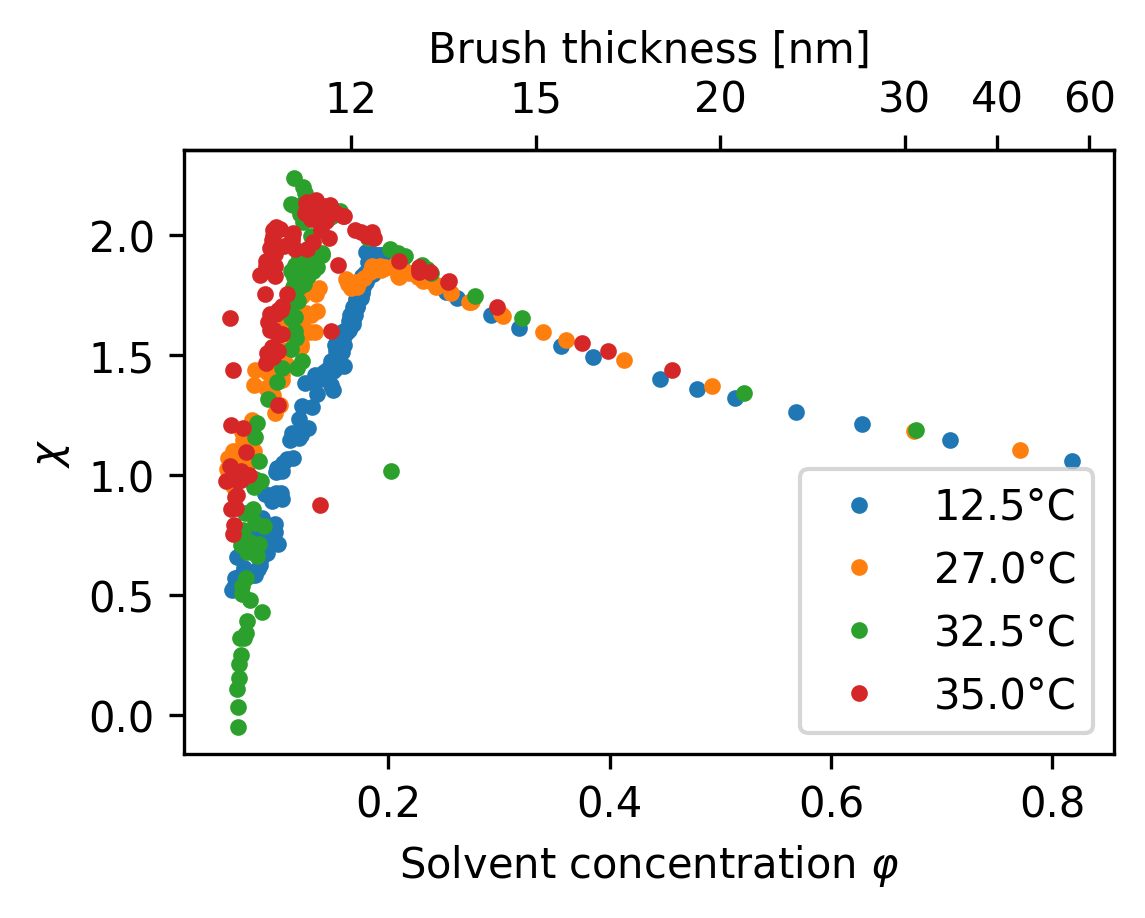}
		\caption{Experimentally determined $\chi(\varphi)$}
		\label{fig:data_phi_vs_chi}
	\end{subfigure}
	\hfill
	\begin{subfigure}[b]{0.45\textwidth}
		\includegraphics[width=\linewidth]{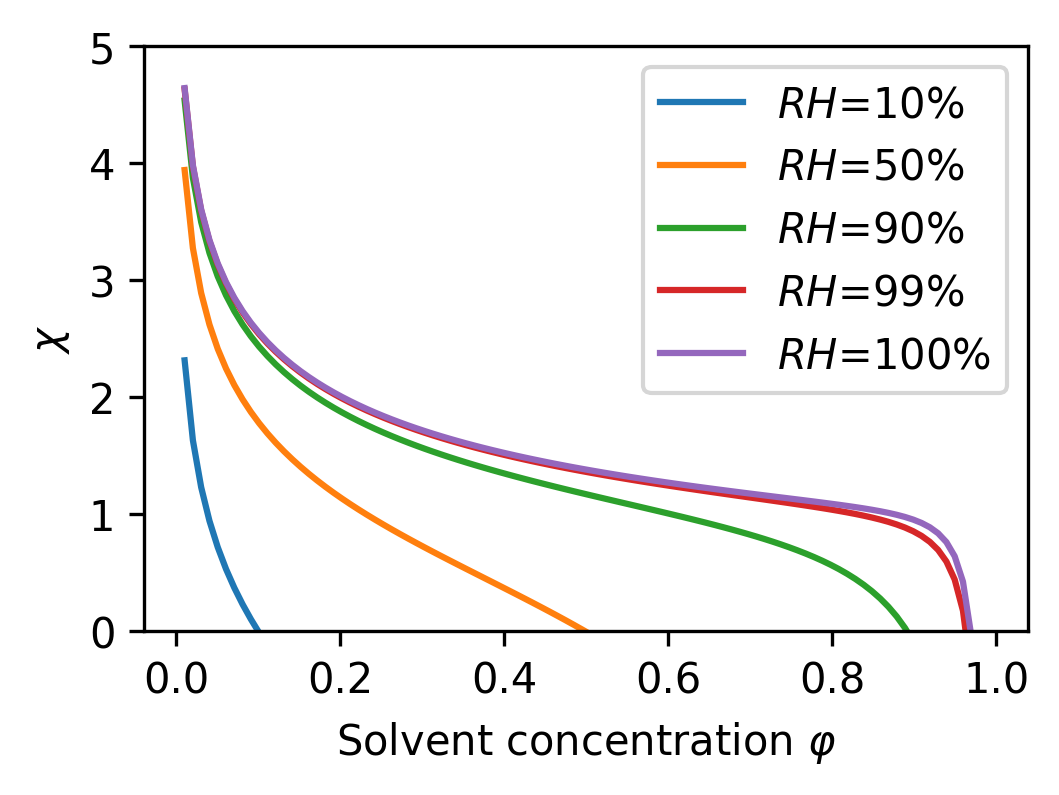}
		\caption{Theoretical sensitivity of swelling to $\chi$}
		\label{fig:phi_vs_chi_fixed_RH}
	\end{subfigure}
	
	\vspace{1em}
	
	\begin{subfigure}[b]{0.45\textwidth}
		\includegraphics[width=\linewidth]{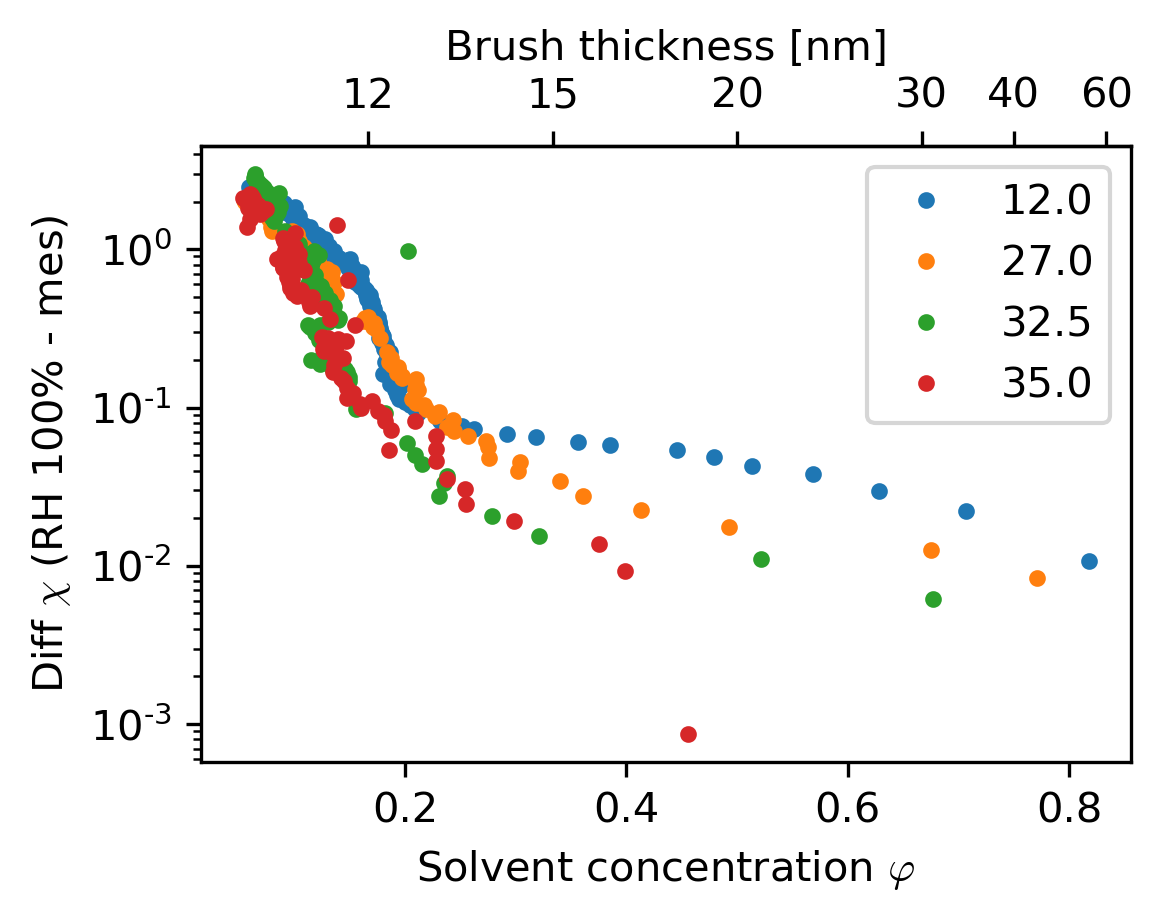}
		\caption{Approach to saturation, $\Delta\chi = \chi_{100\%} - \chi(\varphi)$}
		\label{fig:data_phi_vs_diff_chi}
	\end{subfigure}
	\hfill
	\begin{subfigure}[b]{0.45\textwidth}
		\includegraphics[width=\linewidth]{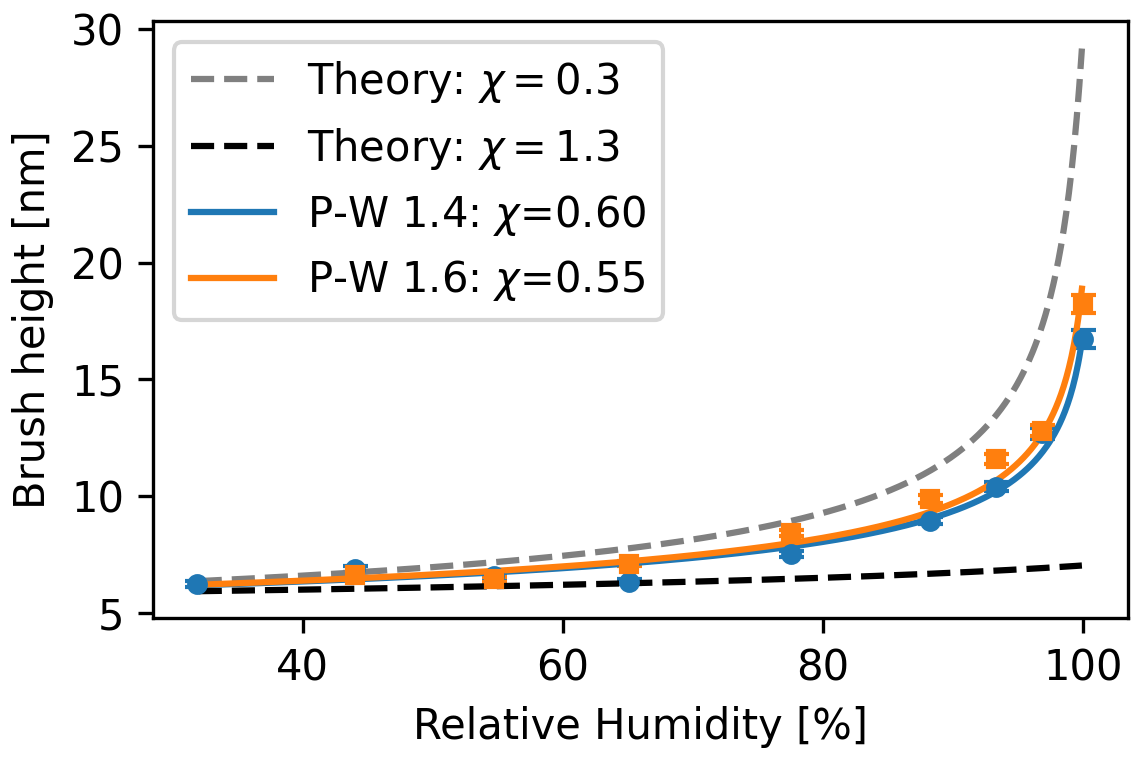}
		\caption{MD Simulation Swelling Isotherms}
		\label{fig:sim_RH_pp_pw}
	\end{subfigure}
	\caption{
	    Detailed analysis of the interaction parameter and comparison with molecular dynamics (MD) simulations.
	    (a) The raw interaction parameter, $\chi(\varphi)$, determined from the experimental isotherms.
	    (b) Theoretical relationship between $\chi$ and $\varphi$ at fixed RH, illustrating the extreme sensitivity of the equilibrium swelling to small changes in $\chi$ near saturation.
	    (c) The difference between the interaction parameter at full saturation ($\chi_{100\%}$) (panel b) and the value of $\chi(\varphi)$ determined from our experimental isotherms (panel a), plotted as $\Delta\chi = \chi_{100\%} - \chi(\varphi)$. This analysis reveals that the measured curves for different temperatures approach saturation via distinct paths.
	    (d) Swelling isotherms from MD simulations, which fail to reproduce the sharp experimental swelling. The solid lines are fits of the standard, constant-$\chi$ mean-field model to the simulation data, yielding $\chi$ values of 0.55 and 0.6.
	}
	\label{fig:Meanfield_modeling_SI}
\end{figure*}

\autoref{fig:data_phi_vs_chi} shows the raw $\chi(\varphi)$ curves determined from the experimental isotherms at four different temperatures. To calculate these values, the experimental data of brush thickness, $h$, versus relative humidity, RH, were first converted into the thermodynamic variables of solvent volume fraction, $\varphi = 1 - h_\mathrm{dry}/h$, and chemical potential, $\mu = k_\mathrm{B}T \ln(\mathrm{RH})$. For each experimental data point $(\varphi, \mu)$, the corresponding value of $\chi(\varphi)$ was then calculated directly by numerically solving Eq.~(3) from the main text. For solvent concentrations $\varphi > 0.3$, the data from different temperatures show an apparent collapse, although subtle differences are present, as detailed below. Below $\varphi \approx 0.3$, this trend is less clear. This regime corresponds to low degrees of swelling, where the calculation of $\chi$ is subject to higher experimental uncertainty due to the larger relative contribution of small errors in the thickness measurement.

To understand how the subtle differences in these $\chi(\varphi)$ curves can lead to the significant differences in maximum swelling, we must examine the system's theoretical sensitivity near saturation. Using the mean-field model, Eq.~\eqref{eq:chi_solution_si}, with a fixed structural parameter ($\rho_g=0.025$), we can plot the theoretical relationship between the interaction parameter $\chi$ and the solvent fraction $\varphi$ required to maintain equilibrium at several constant relative humidities, as shown in \autoref{fig:phi_vs_chi_fixed_RH}. These curves represent all possible equilibrium states for the system, and our experimentally determined $\chi(\varphi)$ dependence corresponds to a specific ``cut'' through this set of curves. The critical feature of this plot is how close the curves for 99\% and 100\%~RH are to each other. This proximity means that a minuscule difference in the $\chi$ parameter can correspond to a dramatic difference in the equilibrium solvent concentration $\varphi$ in this high-humidity regime.

This extreme sensitivity explains why the small differences between the experimental $\chi(\varphi)$ curves are so impactful. To resolve these differences, we analyze the approach to the 100\%~RH state in \autoref{fig:data_phi_vs_diff_chi}. For solvent concentrations $\varphi > 0.3$, these subtle differences follow a clear trend: a higher temperature results in a higher value of $\chi$, consistent with water becoming a poorer solvent for PNiPAAm as the LCST is approached. The plot of the difference, $\Delta\chi = \chi_{100\%} - \chi(\varphi)$, clearly reveals that the curves for different temperatures do not collapse but approach the saturation point via distinct paths. The rank ordering of these paths is consistent with the LCST behavior. It is these subtle, distinct paths of $\chi(\varphi)$, amplified by the system's high sensitivity near saturation, that quantitatively determine the different maximum swelling heights observed at each temperature.

\subsection{\label{app:simulations}Molecular Dynamics Simulations}

As stated in the main text, standard models based on a constant interaction parameter fail to capture the sharp experimental swelling. To illustrate this limitation, we performed coarse-grained molecular dynamics (MD) simulations. The resulting swelling isotherms for two different solvent qualities (varied by the polymer-solvent interaction strength, $\epsilon_\mathrm{ps}$) are shown as symbols in \autoref{fig:sim_RH_pp_pw}.

The simulations successfully reproduce the qualitative behavior observed in our experiments: the degree of swelling is largely insensitive to humidity at lower RH values but increases as saturation is approached. Furthermore, the trend correctly shows that a better solvent (higher $\epsilon_\mathrm{ps}$) leads to more swelling. To quantitatively compare this behavior to the standard theoretical framework, we fitted Eq.~(1) from the main text to the simulation data, assuming a single, constant interaction parameter, $\chi$. As shown by the solid lines in \autoref{fig:sim_RH_pp_pw}, the model provides an excellent fit to the simulated isotherms, yielding constant $\chi$ values of 0.55 and 0.6 for the two solvent qualities.

However, we note that the selection of LJ interaction parameters, and thus the accessible range for a constant $\chi$, was constrained to a narrow, physically-meaningful window. This is a practical limitation of the model: excessively strong polymer-polymer attraction led to unrealistic chain aggregation, while overly strong polymer-water attraction resulted in a non-physical glassy state. The difficulty in finding stable parameters that can span a wide range of solvent qualities with a single, constant $\chi$ is, in itself, an indication of the inadequacy of this approach.

Even within this narrow, valid parameter range, the key result is the fundamental discrepancy with the experiment. Our analysis of the experimental data reveals that $\chi$ is not constant but varies significantly with hydration. Considering only the values at 100\%~RH for different temperatures, the experimental system spans a wide range of effective $\chi$ values from approximately 0.3 (good solvent, low T) to 1.3 (poor solvent, high T). The constant $\chi$ values of 0.55 and 0.6 that successfully describe the simulation are thus unable to describe the broad range and concentration-dependent nature of the interaction parameter required to describe the real system. This demonstrates that a standard coarse-grained model with interactions based on isotropic LJ-potentials is incapable of capturing the highly non-linear response of the PNiPAAm brush near saturation. Rather specialized simulation models would be required to reproduce the LCST behavior of these polymers and approximate this experiment.

\subsection{\label{app:smoothening}Processing of the Swelling Isotherm Data for Derivative Calculation}

Calculating the classical Flory-Huggins parameter, $\chi_\mathrm{class}$, requires taking the derivative of our experimentally determined $\chi(\varphi)$ with respect to the solvent volume fraction, $\varphi$. Directly differentiating the raw experimental data is unreliable, as measurement noise gets amplified, particularly at lower humidities where swelling changes are small. To overcome this, we adopted a hybrid approach to construct the swelling isotherm, $h(\mathrm{RH})$.

First, we fitted a multi-stage analytical function (linear, quadratic, and exponential) to the isotherm at each temperature. This function accurately models the swelling isotherm, capturing both the initial, gradual swelling at lower humidities and the subsequent sharp increase near saturation. However, as the brush approached full saturation, we identified a point, $\mathrm{RH}_\text{switch}$, where the fit began to deviate from the measured data. This switch point varied with temperature but was typically found in the high-humidity regime between 97\% and 99\% RH.

For the final dataset used in our thermodynamic calculations, we used the values from our smooth analytical fit for all humidities below this switch point ($\mathrm{RH} < \mathrm{RH}_\text{switch}$). Above it ($\mathrm{RH} \geq \mathrm{RH}_\text{switch}$), we used the directly measured (binned and averaged) experimental data points. This hybrid method gives us the best of both worlds: it provides a smooth, stable curve for differentiation at lower humidities, while preserving the true experimental behavior in the critical region near saturation where the swelling is sharpest.

\subsection{\label{app:grafting_den}Grafting Density}

The dimensionless grafting density of $\sigma=0.025$ was chosen for the simulation and the mean-field model. This value is not a free fitting parameter but is a physically realistic choice, representative of a typical polymer brush synthesized via the ``grafting-to'' method~\cite{Besford2022}. We validate this choice by demonstrating that it produces a simulated brush whose key dimensionless characteristics are in good agreement with the experimental system.

The primary validation comes from comparing two fundamental dimensionless ratios that characterize the swollen state. First, we consider the chain stretching degree, $h_\mathrm{swollen}/L_c$, which describes the chain conformation relative to its contour length, $L_c$ (the maximum possible length of a fully extended chain). The experimental chain has a contour length of $L_c \approx 135$~nm, giving a stretching degree of $\approx 0.37$. This is in excellent agreement with the simulated chain, which has a contour length of $L_c = 120~d$ and a stretching degree of $\approx 0.27$. Additionally, we compare the swelling ratio, $h_\mathrm{swollen}/h_\mathrm{dry}$, which quantifies the total solvent uptake. Our experimental brush exhibits a swelling ratio of $\approx 4.2$, while the simulation yields a comparable value of $\approx 2.7$.

The good agreement in the stretching degree confirms that our model with $\sigma=0.025$ correctly captures the essential physics of the chain conformation. This value represents a well-balanced choice; while a higher simulated grafting density would further improve the stretching degree match, it would simultaneously hinder swelling due to increased steric repulsion, thereby worsening the agreement for the swelling ratio~\cite{RitsemavanEck2022}.

\begin{figure}
	\centering
	\includegraphics[width=0.8\linewidth]{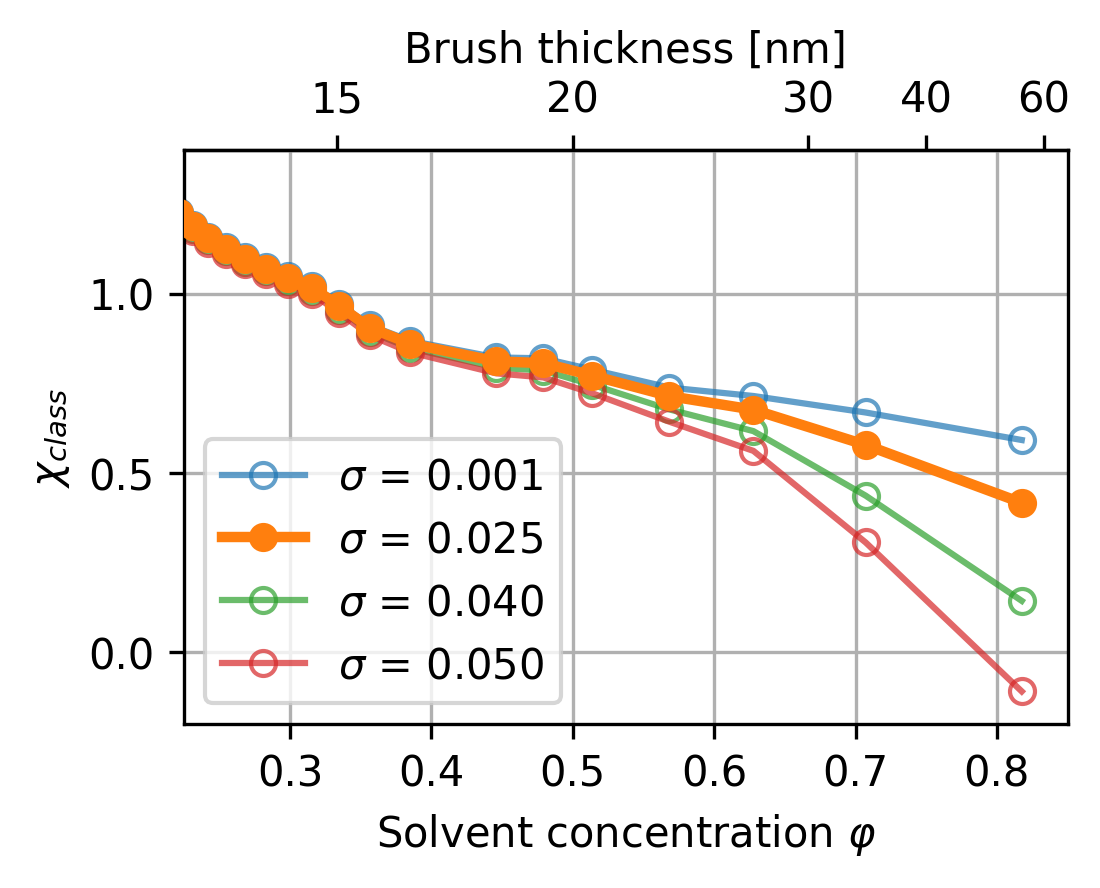}
	\caption{
	    Calculated classical interaction parameter, $\chi_\mathrm{class}$, as a function of solvent fraction, $\varphi$, for different assumed values of the dimensionless grafting density, $\rho_g$. The curves are derived from the 12~$^\circ$C experimental swelling isotherm. A higher assumed $\rho_g$ requires a lower calculated $\chi_\mathrm{class}$ to match the experimental data.
	}
	\label{fig:chi_class_vs_rhog}
\end{figure}

Furthermore, we analyzed the sensitivity of the calculated classical interaction parameter, $\chi_\mathrm{class}$, to the assumed value of $\rho_g$ in the mean-field model. Since the experimental swelling curve is fixed, the choice of $\rho_g$ dictates the value of $\chi_\mathrm{class}$ required to satisfy the model. A larger assumed $\rho_g$ imposes a stronger elastic penalty against swelling, which must be compensated by a stronger driving force for mixing (a lower calculated $\chi_\mathrm{class}$) to reproduce the same experimental data. This relationship is shown in \autoref{fig:chi_class_vs_rhog}, where we plot the calculated $\chi_\mathrm{class}(\varphi)$ curves from the 12~$^\circ$C experimental data for a range of plausible $\rho_g$ values.

Our value of $\rho_g=0.025$ proves to be a physically consistent choice. It correctly predicts that the brush enters a good solvent state ($\chi_\mathrm{class} < 0.5$) when highly swollen, which is the expected behavior for PNiPAAm in cold water. In contrast, assuming a significantly lower grafting density (e.g., $\rho_g=0.01$) would incorrectly suggest the system remains a poor solvent even at full hydration. Conversely, a much higher grafting density would predict an unrealistically good solvent quality across the entire concentration range.

Finally, we confirm that this dimensionless grafting density is consistent with the physical size of the monomers. The relationship $\sigma = \sigma_\mathrm{exp} \times A_m$, using our experimental grafting density of $\sigma_\mathrm{exp} = \SI{0.14}{chains\per\nano\meter\squared}$, implies an effective monomer cross-sectional area of $A_m \approx \SI{0.18}{\nano\meter\squared}$. This corresponds to an effective monomer diameter of $\approx \SI{0.48}{\nano\meter}$, a value that is on the same physical length scale as estimates based on chemical structure ($\approx \SI{0.3}{\nano\meter}$)~\cite{Xue2011}.

\bibliography{Articel}